\documentclass[letterpaper,twocolumn,10pt]{article}
\usepackage{usenix2019,epsfig}
\usepackage{booktabs} 
\usepackage{epstopdf}
\usepackage{enumerate}
\usepackage{graphicx}
\usepackage{pythonhighlight}
\usepackage{authblk}
\newcommand{\eg}{e.g.,}
\newcommand{\ie}{i.e.,}
\usepackage{float}
\usepackage[T1]{fontenc}
\usepackage{listings}
\usepackage{color}
\usepackage{xcolor}
\usepackage{appendix}
\usepackage{hyperref}
\usepackage{placeins}
\usepackage{url}

\renewcommand{\paragraph}[1]{\vspace{5pt}\noindent\textbf{#1}}

\usepackage{adjustbox}
\newcommand{\tabincell}[2]{\begin{tabular}{@{}#1@{}}#2\end{tabular}}

\usepackage{subcaption}

\usepackage{xspace}
\newcommand{\id}{device ID\xspace}
\newcommand{\ID}{\textit{Device ID}\xspace}
\newcommand{\ids}{device IDs\xspace}
\newcommand{\cid}{\textit{Type \uppercase\expandafter{\romannumeral1}}\xspace}
\newcommand{\hid}{\textit{Type \uppercase\expandafter{\romannumeral2}}\xspace}
\newcommand{\phaemph}{\textit{phantom}\xspace}
\newcommand{\pha}{phantom\xspace}
\newcommand{\et}{\textit{et al.}\xspace}
\usepackage{multirow}
\usepackage{url}
\usepackage{amssymb}

\DeclareMathAlphabet{\mathcal}{OMS}{cmsy}{m}{n}

\colorlet{punct}{red!60!black}
\definecolor{background}{HTML}{EEEEEE}
\definecolor{delim}{RGB}{20,105,176}
\colorlet{numb}{magenta!60!black}

\lstdefinelanguage{json}{
    basicstyle=\scriptsize\ttfamily,
    numbers=left,
    numberstyle=\scriptsize,
    stepnumber=1,
    numbersep=8pt,
    showstringspaces=false,
    breaklines=true,
    frame=lines,
    backgroundcolor=\color{background},
    literate=
      {:}{{{\color{punct}{:}}}}{1}
      {,}{{{\color{punct}{,}}}}{1}
      {\{}{{{\color{delim}{\{}}}}{1}
      {\}}{{{\color{delim}{\}}}}}{1}
      {[}{{{\color{delim}{[}}}}{1}
      {]}{{{\color{delim}{]}}}}{1},
}

\definecolor{codegreen}{rgb}{0,0.6,0}
\definecolor{codegray}{rgb}{0.5,0.5,0.5}
\definecolor{codepurple}{rgb}{0.58,0,0.82}
\definecolor{backcolour}{rgb}{0.95,0.95,0.88}

\lstdefinestyle{mystyle}{
    backgroundcolor=\color{backcolour},
    commentstyle=\color{codegreen},
    keywordstyle=\color{magenta},
    numberstyle=\tiny\color{codegray},
    stringstyle=\color{codepurple},
    basicstyle=\scriptsize\ttfamily,
    breaklines=true,
    captionpos=b,
    keepspaces=true,
    numbers=none,
    numbersep=5pt,
    showspaces=false,
    showstringspaces=false,
    showtabs=false,
    tabsize=2
}
\usepackage{tikz}

\begin{document}


\title{\Large \bf Discovering and Understanding the Security Hazards in the Interactions between IoT Devices, Mobile Apps, and Clouds on Smart Home Platforms}
\author[1]{Wei Zhou}
\author[2,1]{Yan Jia}
\author[2,1]{Yao Yao}
\author[2,1]{Lipeng Zhu}
\author[3]{\\Le Guan}
\author[2,1]{Yuhang Mao}
\author[4]{Peng Liu}
\author[ ]{Yuqing Zhang$^{1,2,5}$\thanks{Corresponding author: zhangyq@nipc.org.cn}}
\affil[1]{\textit{National Computer Network Intrusion Protection Center, University of Chinese Academy of Sciences, China}}
\affil[2]{\textit{School of Cyber Engineering, Xidian University, China}}
\affil[3]{\textit{Department of Computer Science, University of Georgia, USA}}
\affil[4]{\textit{College of Information Sciences and Technology, The Pennsylvania State University, USA}}
\affil[5]{\textit{State Key Laboratory of Information Security, Institute of Information Engineering, Chinese Academy of Sciences, China}}
\renewcommand*{\Affilfont}{\normalsize\it} 
\renewcommand\Authands{ and }

\maketitle


\pagestyle{empty}
\thispagestyle{empty}

\vspace*{-9mm}
\subsection*{Abstract}
\vspace*{-1mm}
A smart home connects tens of home devices to the Internet,
where an IoT cloud runs various home automation applications. 
While bringing unprecedented convenience and accessibility, it
also introduces various security hazards to users.
Prior research studied smart home security from several aspects.
However, we found that the complexity of the interactions among the participating entities (i.e., devices, IoT clouds, and mobile apps) has not yet been systematically investigated.
In this work, we conducted an in-depth analysis of
five widely-used smart home platforms.
Combining firmware analysis,
network traffic interception, and black-box testing,
we reverse-engineered the details of the interactions among the participating entities. Based on the details, we
inferred three legitimate state transition diagrams for the
three entities, respectively. Using these state machines
as a reference model, we identified a set of unexpected state transitions. To confirm and trigger the unexpected state transitions, we implemented a set of phantom devices to mimic a real device. By instructing 
the phantom devices to intervene in the normal entity-entity interactions, we have discovered several new vulnerabilities
and a spectrum of attacks against real-world smart home platforms. 


\vspace*{-2.5mm}
\section{Introduction}
\label{sec:intro}
\vspace*{-2mm}

With the development of the Internet of Things (IoT), smart home technology has become widely used in many applications including safety and security~\cite{homesafety}, 
home appliances~\cite{homeappliance}, home healthcare~\cite{healthcare}, etc.
According to Statista research, more than 45 million smart home devices were installed in 2018, and the annual growth rate of home automation is 22\%~\cite{SH2018}.
To manage the ever-increasing number of diverse smart home devices in a consolidated way,
many companies have proposed their smart home platforms (\eg~Samsung SmartThings~\cite{SmartThings}).
With IoT clouds playing a central role in building smart homes,  
real-world smart home platforms essentially engage \textbf{three (kinds of) entities} that interact with each other: 
an IoT cloud, smart home devices and a mobile app.
Briefly speaking, the mobile app provides users with an interface to facilitate the initial setup of devices including WiFi provision.
After getting Internet access, each device negotiates its login credential(s) with the IoT cloud. 
In this way, it can build a connection with the IoT cloud to routinely report its status and execute the received remote control commands, which are usually generated by certain home automation applications running in the IoT cloud. 
At the same time, the mobile app is able
to monitor and control each device through the IoT cloud.

While bringing substantial convenience to our lives, smart home technology also introduces potential security hazards.
Since smart home devices directly process user-generated data, once compromised, they could introduce serious consequences. For example, user privacy can be harmed~\cite{thermostat,newman2018turning}; property can be destructed~\cite{Fereidooni2017Breaking,Ransomware}; life safety and psychological health are also threatened~\cite{HeartAttack,Valente2017Security}.
Imagine a smart home which is programmed in such a way that whenever  the home temperature rises to a given threshold,
the windows will be automatically opened.
If an attacker obtains access to a smart heater, he could easily
break into the home by keeping the heater at the highest temperature~\cite{ding2018safety}.

Although an increasing number of research studies have focused on smart home security, we found that existing research on the insecurity of interactions (e.g. inter-operations) in smart home platforms is still quite limited.  
First, the existing studies usually focus on individual parts of smart home platforms. 
For instance, there are studies disclosing the security problems with   
device firmware~\cite{Notra2014An,Ling2017Security,thermostat}, communication protocols~\cite{HackingZWave,Ronen2017IoT,Goyal2016Mind}, and home automation applications~\cite{Fernandes2016Security,lee2017fact,celik2018sensitive}.
Focusing on individual parts, the revealed vulnerabilities have little to do with the interactions among the three entities engaged in a smart home platform.  
Second, the existing studies seem to pay most attention to classic security issues such as privacy protection~\cite{Earlence2016FlowFence,yu2018pinto}, authentication~\cite{Notra2014An,Ling2017Security} 
and permission models~\cite{fernandes2018decentralized,Fernandes2016Security,lee2017fact}, 
and leave the potential risks of the entity-entity interactions largely uninvestigated.

Third, one kind of interaction in the Samsung SmartThings platform has recently been studied in  ~\cite{celik2018soteria,celik2018sensitive}, finding that the interaction between multiple home automation applications (i.e. IoT apps) can lead to unsafe device states. 
While this finding is inspiring, other essential kinds of interactions in smart home platforms have not yet been (systematically) studied in the literature. 
In this paper, when we use the word ``interactions'', we are specifically referring to the inter-operations involved among the aforementioned three entities, with a focus on high-level pairing of devices, handshaking between IoT clouds and devices, etc.

To systematically discover and understand the security hazards in the interactions 
involved in smart home platforms, 
we have analyzed several widely-used smart home platforms
and 
conducted the following investigations in this work. 
First, because all the communications among the three entities are encrypted, 
we combined several techniques including
firmware reverse-engineering and man-in-the-middle (MITM) monitoring (to break SSL)
to work out the details of the interactions among the three entities.
Second, based on the interactions among the three entities, we inferred three legitimate state transition diagrams for the three entities, respectively. Using the inferred state machines as a reference model, 
we identified unexpected state transitions in several widely-used smart home platforms. 
Finally, 
to confirm and trigger unexpected state transitions,
we implemented a \phaemph device that mimics a real device.
A phantom device is essentially a computer program.
By instructing the phantom device to intervene in the normal interactions among legitimate smart home devices, IoT clouds, and a mobile app, 
we have identified several new vulnerabilities and attacks in major smart home platforms.



In summary, the main contributions are as follows:
\vspace*{-1.5mm}

\paragraph{New insights are provided:}
(a) Real-world smart home platforms do not strictly guard the validity of the involved state transitions.
For example, we found that an IoT cloud can accept some device requests without checking whether such a request should be allowed or not in its current state.
(b) The three entities can sometimes stay
in unexpected state combinations, which brings potential risk.
(c) IoT clouds do not always perform adequate authorization checks on interaction requests. 
We found that an IoT cloud sometimes simply accepts and executes sensitive device-side commands without any permission checking. 
(d) By carefully constructing attacks that exploit a particular combination of the above security flaws, we showed that serious new security hazards can occur. This {\em new} finding proves that high risk attacks are rarely caused by a single factor. 
Accordingly, stake holders should conduct integrated insecurity analysis on interactions among the three entities.

\paragraph{New hazards are discovered:}
(a) An adversary can remotely replace a victim's real device with a non-existing phantom device under his control. As a result,
all the control commands from the victim user are exposed to the phantom device and further to the adversary,
leading to privacy breaches.   
The adversary can also leverage the phantom device to manipulate the data to be sent to the victim user, 
thus deceiving or misleading the victim user.
(b) An adversary can remotely take over a device.
As a result, he can harvest the sensor readings to monitor the victim's home or even manipulate the smart home devices, 
causing data breaches and endangering the victim.
(c) An adversary can remotely unbind an authorised user through a phantom device.
As a result, the user can no longer control the device with his account.
(d) An adversary can leverage a phantom device to mislead an IoT cloud and occupy the identity of a real device before the device is sold. When a consumer buys the device, he cannot bind the device with his account.
(e) An adversary can utilize a phantom device to automatically send update requests
to an IoT cloud to steal various proprietary firmware on a large scale. 

The newly discovered hazards have significantly enlarged the previously-known attack surface of smart home platforms; 
they also provide essential new understandings about the security and privacy hazards in smart homes.


\emph{Responsible Disclosure.}
All the vulnerabilities described in this paper have been reported to the corresponding vendors, and they 
have confirmed our disclosures.
We have shared the technical details with the vendors. And most of the vulnerabilities have been fixed by them.

\vspace*{-2mm}
\section{Background}
\label{background}
\vspace*{-1.5mm}

\subsection{Terminology}
\label{sec:terms}
To make the presentation more clear,
we first define several key terminologies.

\paragraph{Device ID.} 
The \emph{Device ID} of an IoT device
uniquely identifies the device. 
Since device IDs are used to authenticate a device, they should be kept secret at all time. 
The attacks discovered in this study create fake IoT devices by
occupying the device ID of a real victim device. 

\paragraph{Identity Information.}
By ``identity information,'' we mean the information items whose values are used to generate (i.e. calculate) a device ID. 
A typical use case of identity information is as follows: a device first provides the IoT cloud it belongs to with
its identity information, then the cloud generates and returns the corresponding device ID to the device. 
Frequently used identity information includes
MAC address and device model.
Since device IDs should be kept secret, identity information should also be kept secret.
Unfortunately, we found that this rarely holds in practice and attackers can easily obtain device identity information.

\paragraph{Legitimacy Information.}
By ``legitimacy information,'' we mean the information items whose values are used to conduct certain legitimacy checking of a device, but are not used to generate any device IDs. 
We found that such information can also be easily
acquired by hackers. 

\paragraph{Phantom Devices.} 
A phantom device is used by us to analyze the 
smart home interactions and to launch attacks.
It is essentially a computer program that is instructed to intervene in the normal
interactions among legitimate smart home devices, 
IoT clouds, and mobile apps.



\begin{figure}[t]
\includegraphics[width=\columnwidth]{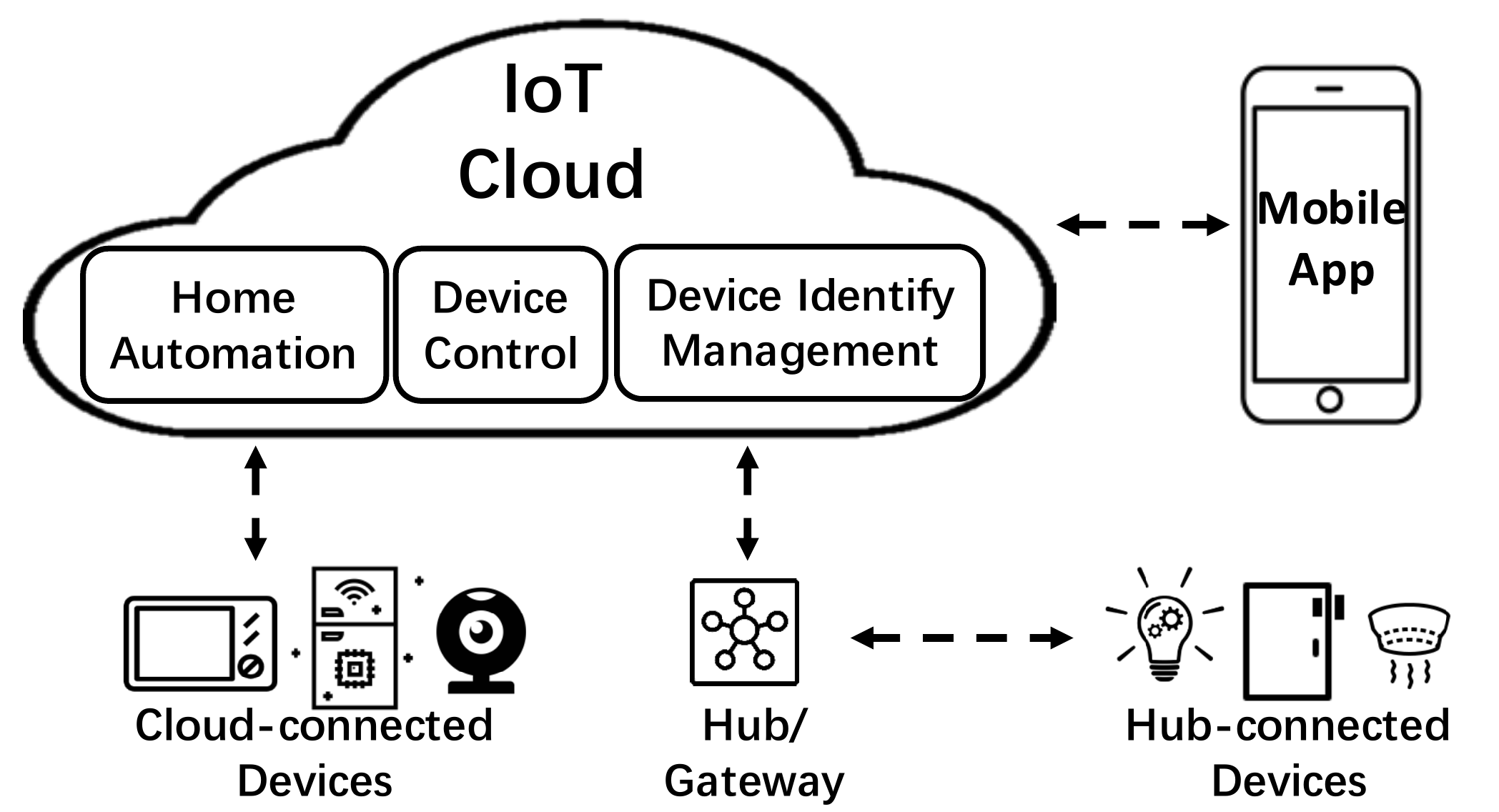}
\caption{Smart Home Platform Architecture}
\label{fig:platform}
\vspace*{-5mm}
\end{figure}

\vspace*{-1.5mm}
\subsection{Overview of Smart Home Platforms}
\vspace*{-1.5mm} 
\label{sec:overview_arch}


The architecture shown in Figure~\ref{fig:platform} is widely adopted by major cloud-based smart home platforms. 
There are three key entities interacting with each other
on a smart home platform: the IoT devices, the mobile app, and the IoT cloud. 

The IoT cloud is the brain of a smart home platform.
It is usually responsible for three kinds of services, denoted as Device Identify Management, Device Control, and Home Automation as shown in Figure~\ref{fig:platform}. 
First, in order to ensure that only the device owner and delegated users have access to a  device,
the device identify management service needs to 
maintain a one-to-one mapping between the owner's account and the device.
This binding happens at the time when the device
is firstly deployed.
As soon as the device is undeployed, the binding relationship should be revoked.
Second, in order to allow authorized users to remotely control a device,
the device control service serves as a ``proxy'' when users send remote commands to the device.
Lastly, most smart home platforms provide home automation services, in which users can customize automation rules that define the interoperability behaviors of smart home devices.
For example, a home owner can craft an automation rule that turns on 
the air conditioner if the indoor temperature goes above 70$^\circ$F.
When a thermometer detects that the temperature exceeds the specified threshold, it sends the event to an in-cloud home automation application, which then sends a command to turn on the air conditioner. 

The second type of entities in a smart home platform are IoT devices.
IoT devices are equipped with embedded sensors and actuators  
that interact with the physical world and send sensor readings to an IoT cloud.  
There are two typical mechanisms for devices to connect to an IoT cloud. 
(a) WiFi-enabled devices can connect to the Internet
and thus directly communicate with the IoT cloud.
We call these devices \textit{cloud-connected devices}.
(b) Energy-economic devices are not equipped with a WiFi interface to directly interact with the IoT cloud. Instead, they first connect
to a hub/gateway using energy-efficient protocols such as Z-Wave and ZigBee. 
Then the hub connects to the IoT cloud on behalf of the IoT devices. 
We call the devices connected to a hub as \textit{hub-connected devices}.
It is worth noting that the hub itself is one kind of cloud-connected device.
Some platforms support both 
cloud-connected and hub-connected devices,
while some only support one kind.
The third kind of entities on a smart home platform are mobile apps.  They provide users with an interface for
device management (e.g., binding a device with its owner's account) and
customization of the in-cloud home automation services.

\paragraph{Deployment.}  To standardize and simplify the deployment of IoT services, smart home platform providers often provide collaborating partners with software development kits (SDKs). With SDKs, the adopting manufacturers
only need to focus on device-specific initialization procedures and core application logic.  
Features such as over-the-air (OTA) update are
also integrated in the SDKs.

\begin{figure*}[t]
\centering
\subcaptionbox{State Machine of an IoT Cloud\label{fig:s1}}{\includegraphics[width=0.33\textwidth]{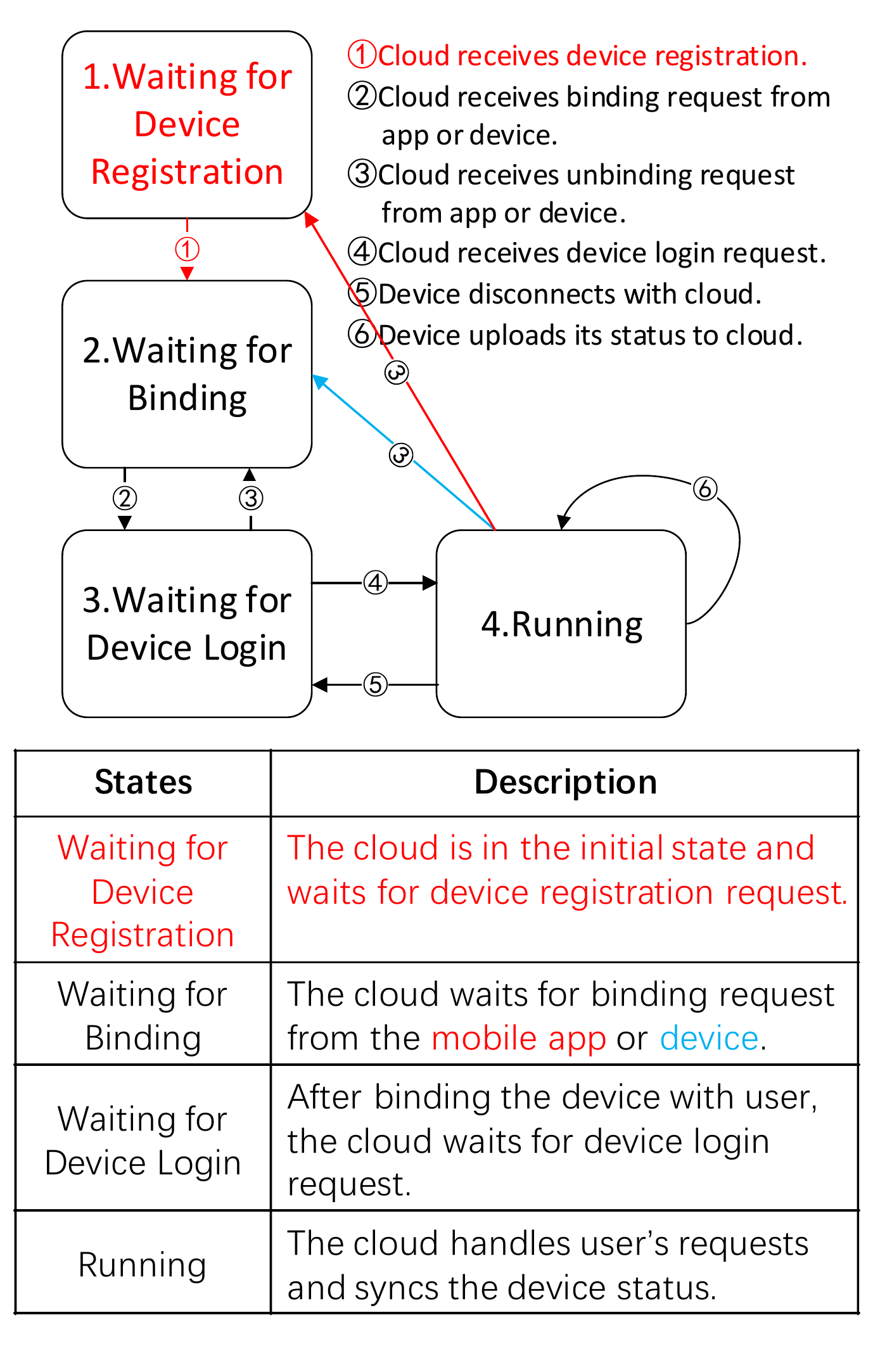}}%
~
\subcaptionbox{State Machine of a Device \label{fig:s2}}{\includegraphics[width=0.33\textwidth]{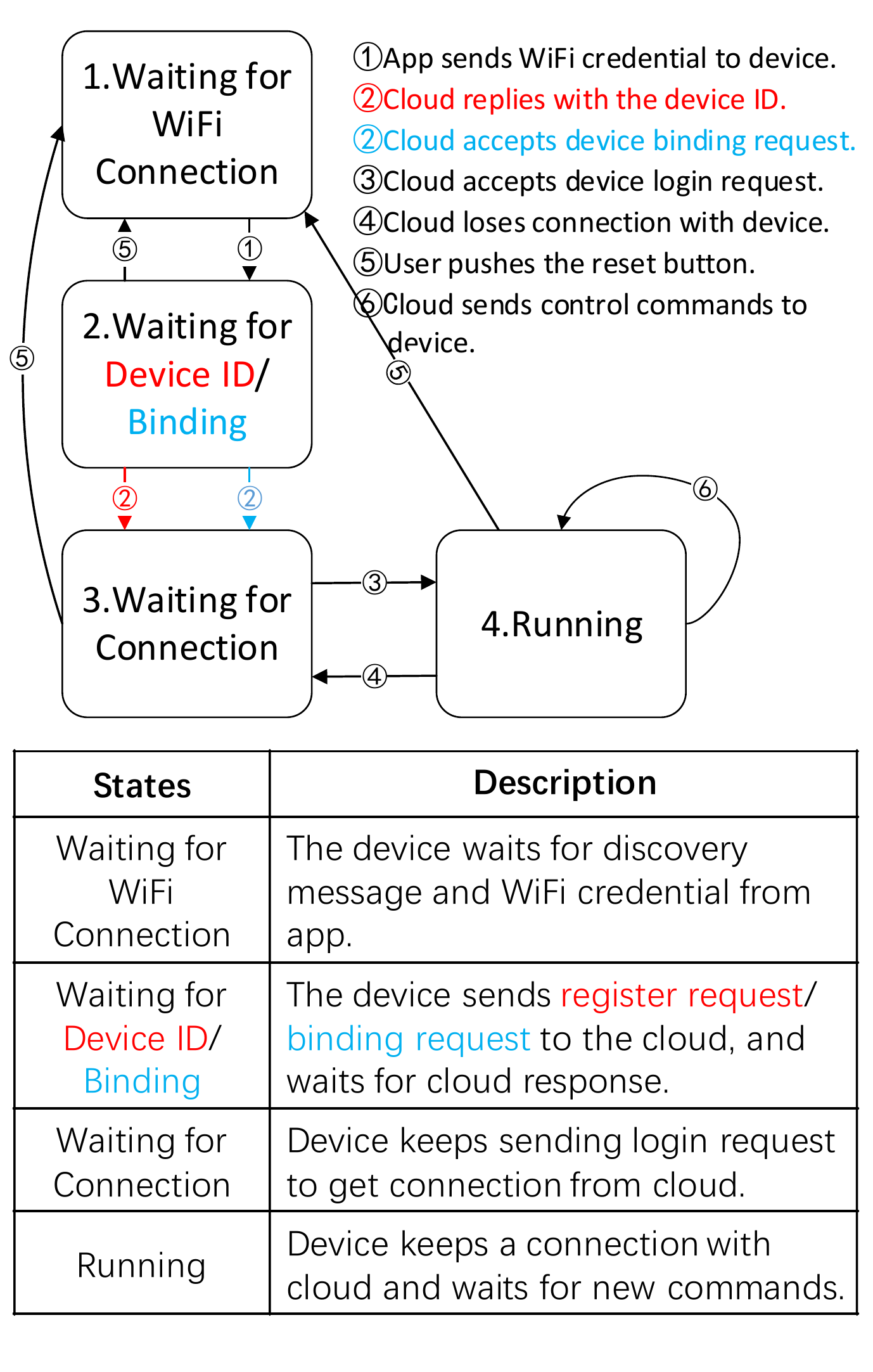}}%
~
\subcaptionbox{State Machine of a Mobile App\label{fig:s3}}{\includegraphics[width=0.338\textwidth]{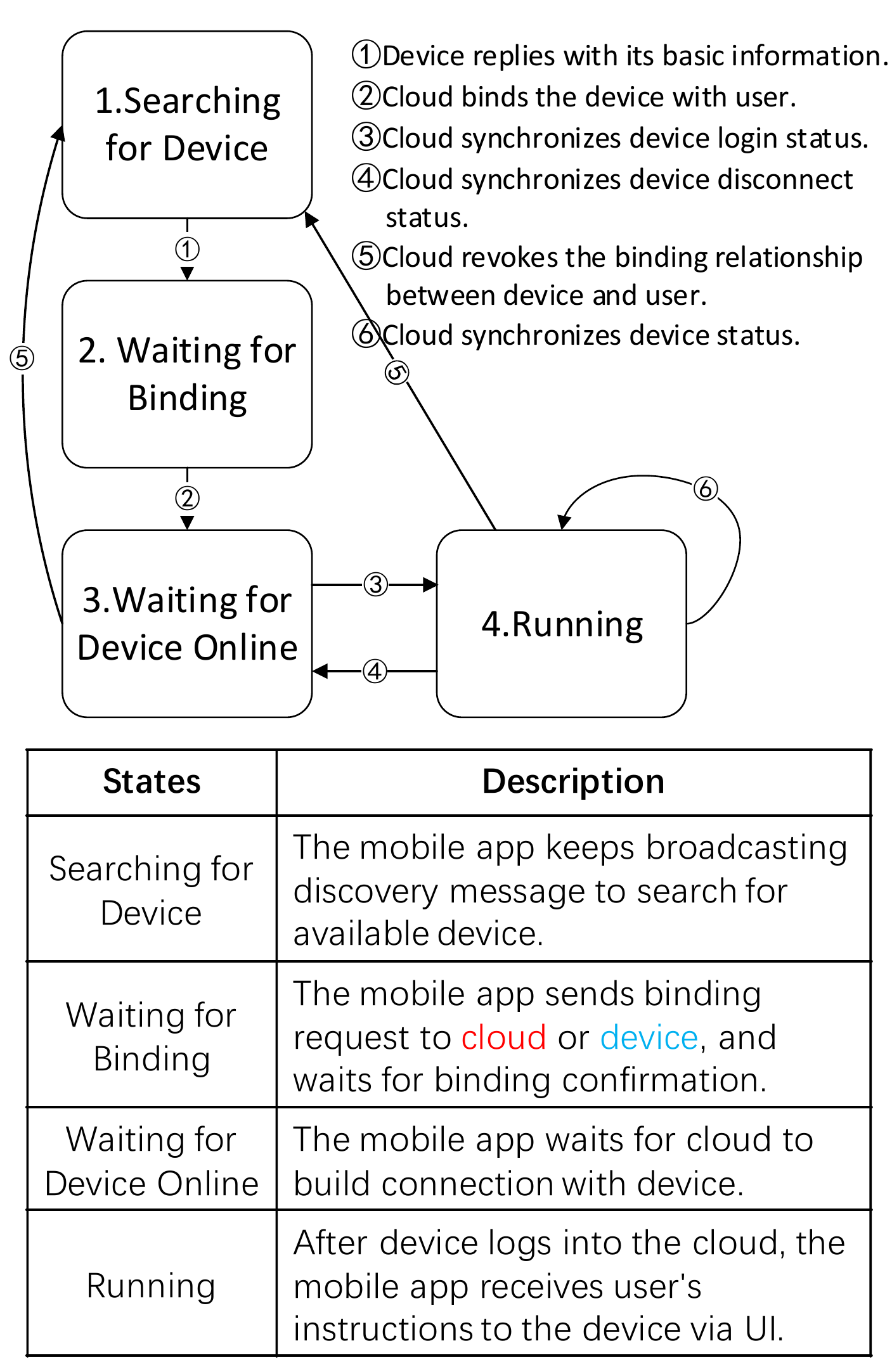}}%
\flushleft
\footnotesize{
Note: The states and transitions specific to
\cid platforms are shown in red; the states and transitions specific to
\hid platforms are shown in blue; and the shared states/transitions are in black.
}
\caption{High-Level State Machines of the Three Entities}
\label{fig:state}

\vspace*{-3mm}
\end{figure*}

\begin{table}[t]
  \caption{Two Types of Smart Home Platforms and Their Differences}
  \vspace*{-4mm}
  \begin{center}
  \begin{adjustbox}{max width=\columnwidth}
  \begin{tabular}{c|ccc}
    \toprule
    &\textbf{Platform}&\textbf{Device Registration}&\textbf{Device Binding/Unbinding}\\
    \midrule
    \textbf{\cid Platform}&Alink, Joylink&\tabincell{c}{\ID\\Generated by \textbf{Cloud}}&\tabincell{c}{Authorization Checking\\Performed by \textbf{Cloud}}
    \cr\hline\hline
    \textbf{\hid Platform}&\tabincell{c}{SmartThings\\KASA, MIJIA}& Skipped &\tabincell{c}{Authorization Checking\\Performed by \textbf{Device}}\\
  \bottomrule
\end{tabular}
\end{adjustbox}
\end{center} 
\label{tab:cate}
\vspace*{-6mm}
\end{table}

\vspace*{-2mm}
\subsection{Overview of the Interactions on Smart Home Platforms}
\label{sec:interaction}
\vspace*{-1.5mm}

In this section, we depict the interactions among the three entities during the life-cycle of a smart home device from the viewpoint of a consumer (rather than a manufacturer, supplier or retailer). To make the description more clear, the description focuses on cloud-connected devices. Hub-connected devices follow a similar model except that they use a hub as an intermediate node.

\cid vs. \hid: 
Depending on how the device ID of a device is generated, we classify the smart home platforms investigated in this work into two types. 
How a device ID is generated further influences device registration and device binding/unbinding.
In Table~\ref{tab:cate}, we list the investigated platforms for each type and
the key differences between the two types.
These differences lead to different attack preconditions in attacking a smart home platform.

In the following, 
we try to abstract an interaction model for \cid and \hid smart home platforms. 
When there is a difference between the two types, 
we also explicitly report it.
In Appendix~\ref{sec:appendix3}, the complete interaction diagrams for all the studied platforms are given.





\paragraph{1. Device Discovery:}
The life-cycle of a newly bought smart home device starts with device discovery. 
After the home owner (i.e. the customer) clicks the ``Add Device'' button on the designated mobile app,  
the app establishes a local connection with the device through  broadcasting a discovery message or through the device's access point. 
Then, the device reports its basic information, such as its MAC address and device model, to the mobile app. 



\paragraph{2. WiFi Provisioning:}
To access the Internet, the IoT device needs to join the same LAN as the mobile app. To obtain WiFi credentials, there are several mechanisms such as
Access Point Mode~\cite{Chang2015Design}, WiFi direct~\cite{WiFidirect} and SmartConfig~\cite{SmartConfig}.

\label{sec:deviceid}
\paragraph{3. Device Registration:}
After device registration, the IoT device is given a unique \id. 
Different types of platforms provide device IDs in different ways. 
For \cid platforms, the device sends its device identity information to the IoT cloud it belongs to. 
The cloud then generates a \id and returns it
to the device. 
The device then writes the device ID to its persistent storage.
The cloud also keeps a record of the \id for future authentication.
For Type II platforms, 
the device's device ID is generated by the device's platform beforehand and hard-coded into the device during fabrication.
Therefore, device registration is skipped. 


\paragraph{4. Device Binding:}
The IoT cloud binds the device's device ID with the user account of the home owner. As a result, only authorized users can access the device via the cloud.
The two types of smart home platforms adopt 
different device binding methods.
For \cid platforms, the binding request is directly sent by the mobile app 
to the IoT cloud, which is responsible for maintaining the binding information and
performing the permission checks (\ie~whether a user account should have access to a device).
For \hid platforms, the mobile app first sends
the account information 
to the device.
The device, having the \id and the user account, issues the binding request to the IoT cloud.
It is worth noting that here the cloud unconditionally accepts the binding request from the device. 
This design is based on the natural assumption that the customer who physically owns a device
should have full control over it. 

\label{sec:interactionprocess}



\paragraph{5. Device Login:}
The device uses its \id to log into the IoT cloud.
Then it establishes a connection with the cloud to keep the status updated 
and ready itself to execute remote commands.
In addition, when a device loses connection with the cloud for a long time,
it tries to re-login automatically.

\paragraph{6. Device in Use:}
After successful registration and login, the device performs designed functions.
Specifically, the home owner can monitor the real-time status of the device and explicitly 
control the device locally or remotely via the ``control panel'' on the mobile app. 

\paragraph{7. Device Unbinding \& Device Reset:}
When the home owner no longer uses the device,
she can unbind or reset it.
When the user requests for unbinding,
for \cid platforms, the cloud directly erases the binding information.
For \hid platforms, however, since the binding information is also stored on the device locally,
one additional command is sent from the cloud to the device to erase the binding information. 


In the life-cycle of an IoT device, although most of the time is spent 
in the sixth phase, \ie~the device-in-use phase,
the interactions that occur during the other phases are the most complex and critical.
Any oversight in these phases could lead
to serious security problems and harm the normal use of devices.

\vspace*{-3.5mm}
\subsection{State Transitions}
\label{sec:state_transition}
\vspace*{-2mm}

In this subsection, we describe the state transitions inferred from our analysis of five widely-used smart home platforms (i.e.,  
Samsung SmartThings~\cite{SmartThings},
TP-LINK KASA~\cite{TP}, XiaoMi MIJIA~\cite{XiaoMi}, Ali Alink~\cite{Ali}, and JD Joylink~\cite{JD}).  
A smart home platform is a special kind of distributed system. In this viewpoint, the aforementioned interactions among IoT devices, mobile apps, and IoT clouds unavoidably cause state transitions. 
Based on our analysis of the aforementioned five platforms,    
we infer three state transition diagrams for the three entities, respectively.  
The three state machines are shown in Figure~\ref{fig:state}. 
In each sub-figure, an interaction with other entities (denoted by an edge) causes a state transition. 
The definitions for each state and each transition are annotated in the corresponding sub-figure.   
Note that 
\cid and \hid platforms behave slightly differently and
we highlight the differences using different colors. 
Specifically, the states and transitions specific to
\cid platforms are shown in red while the states and transitions specific to
\hid platforms are shown in blue.
The shared states and transitions are shown in black. 
In addition, since \hid platforms use hard-coded \id, state 1 is absent in the state machine of an IoT cloud.

\paragraph{State Correlation.}
The three state machines are closely related to one another.
Whenever an interaction takes place, 
the three entities as a whole may transfer from one 3-tuple state combination (i.e., the current state of the IoT cloud, the current state of the device, and the current state of the mobile app) to another 3-tuple.     
We identify all the legitimate 3-tuple state combinations and show them in the table presented in Appendix~\ref{sec:appendix1}.
If a 3-tuple does not appear in the table, the corresponding state combination is illegal and might be exploited.
To avoid the potential attacks,
the three entities should always stay in a legitimate state combination. 
Unfortunately, we found that none of the investigated smart home platforms strictly maintain a three-entity state machine. 

\vspace*{-3mm}
\subsection{Scope of Empirical Vulnerability Analysis}
\label{sec:scope}
\vspace*{-1.5mm}



Real-world cloud-based smart home platforms can be classified into two categories.  
The first category is the platforms  {\em dedicated} to building a smart home (e.g., Samsung SmartThings~\cite{SmartThings}). 
The second category is general-purpose IoT platforms (e.g., Amazon Web Services IoT~\cite{AWSIoTCore}) which could be customized for smart home usage.
Since smart home platforms of the second category usually differ from each other
in terms of device management and interaction, 
we leave studying common security issues with them as our future work.  In addition, smart home platforms that are not cloud-based,  such as HomeKit~\cite{HomeKit} and
HomeAssistant~\cite{HA}, are out of the scope of this study, 
although we will discuss the implications of our research findings to platforms that are not cloud-based    
in Section~\ref{sec:non-cloud}. 


In this work, we focus on five leading cloud-based smart home platforms. As mentioned earlier, they are 
SmartThings, KASA, MIJIA, Alink, and Joylink. 
To attract
more cooperative manufacturers, some smart home platform
providers such as Samsung, JD, and Ali make their platforms open and even open-source the corresponding
device-side SDKs. 
Thus, the collaborative manufacturers can easily follow the documentation and assemble platform compliant devices through proper use of the SDKs. 
Over 200 well-known smart home device manufacturers (e.g., Philips, ECOVACS, and Media)
are actually fabricating products running on these platforms~\cite{Alimarket,JDmarket}.


Some other cloud-based smart home platform providers, including
 TP-LINK and XiaoMi, adopt a closed ``ecosystem''. They  
fabricate smart home devices by themselves. 
On North America and Europe markets, 
TP-LINK's smart home devices, such as smart WiFi plugs and smart LED bulbs, rank in the top 10 in the category of home improvement on Amazon~\cite{TPMarket}.  
XiaoMi is the world's largest intelligent smart home device manufacturer. 
More than 85 million smart home devices have been sold under 
the brand of XiaoMi all over the world~\cite{XiaoMiMarket}, especially in Asia-Pacific~\cite{MijiaMarket}.




\vspace*{-2mm}
\section {Threat Model and Feasibility Assessment}

\vspace*{-2mm}
\subsection{Threat Model}
\vspace*{-2mm}

In contrast to 
network-based exploits (\eg~MITM) and firmware-based reverse-engineering,
the adversary in our threat model seeks to exploit 
the design flaws in the interactions among
the three entities. 
Therefore, we \emph{do not} assume any forms of software bugs or protocol vulnerabilities.
The targets of the attack are cloud-connected devices which directly communicate with IoT clouds.
The adversary's goal is to take control of the device or to monitor/manipulate
the data collected/generated by the device.




\label{sec:threat}
We \emph{do} assume that the adversary has the capability
to collect necessary information, including
device \emph{identity information} and \emph{legitimacy information}.
For different platforms, the difficulty levels of collecting
these information items differ.
For example, for \hid platforms, we assume the adversary has
local access to the victim device beforehand,
whereas for \cid platforms, we do not have such an assumption.
As a result, the discovered exploits may 
exhibit different levels of feasibility depending on 
which type of platform is being attacked, who is the platform provider, etc.

In the following, for both types of
platforms, we analyze the feasibility of obtaining 
these information items case by case.
The reason is that different platforms may designate and use individual information items in different ways.
Correspondingly, the adversary faces different challenges in collecting them.
\vspace*{-3mm}
\subsection{Prerequisites and Feasibility Assessment}
\vspace*{-1.5mm}

\label{sec:fea}

\begin{table}[t]
  \caption{Device Identity/Legitimacy Information}
  \begin{adjustbox}{max width=\columnwidth}
  \begin{tabular}{c|ccc}
    \toprule
    &Platform&Identity Info &Legitimacy Info\\
    \midrule
 \multirow{2}{*}{\textbf{\cid Platform}}&Alink&\tabincell{c}{MAC (G), CID (P),
 \\Device Model (P)}&Key (P), Sign (P)\\
                &Joylink&\tabincell{c}{MAC (G), SN (H),\\Device Model (P)}&NULL
                \cr\hline\hline
 \multirow{3}{*}{\textbf{\hid Platform}}&SmartThings&Device ID (H)&NULL\\
                    &KASA&Device ID (H)&MAC (G), hwID (P)\\
                    &MIJIA&Device ID (H)&TagKey (H)\\
  \bottomrule
\end{tabular} 
\end{adjustbox}
\label{tab:info}
\flushleft
\scriptsize{
P: Public information \hfill G: Guessable information \hfill H: Hard-coded information}
\vspace*{-4mm}
\end{table} 

In this subsection, we describe the 
specific identity and legitimacy information items the adversary has to obtain, and evaluate the feasibility of obtaining them in practice.

As mentioned earlier, a device is identified by a unique \id.
In essence, the discovered exploits fake a \pha device by using
the victim device's device ID.
Thus, the adversary needs to get the device ID of the victim device.  
For \cid platforms, given that the device ID is determined
solely by the victim device's identity information, 
the adversary only needs to collect all the identity information.  

For \hid platforms, the device ID is hard-coded in the victim device.
So the adversary has to have local access
to the victim device (\eg~connect to the same LAN or physically possess the device) to obtain the device ID. 
Although this seems to be a strong assumption, we note that 
once the hard-coded information is leaked, the victim device becomes remotely vulnerable forever.  

Furthermore, some platforms use pre-configured legitimacy information (e.g., a key) as additional authentication requirements.


Depending on the way to
obtain a particular identity/legitimacy information item,
we classify these information items into three categories:
public information (\textbf{P}), guessable information (\textbf{G}) and hard-coded information (\textbf{H}). 
For each platform investigated in this study, we list the needed identity/legitimacy information items plus their categories in Table~\ref{tab:info}.
Note that the same information item may be used differently.
For example, \texttt{MAC} addresses 
are used by Alink devices as identity information, but are used by KASA devices as legitimacy information.


\paragraph{Public information} is the easiest to obtain.
Information items in this category are often publicly available or can be easily inferred.
For example, 
device model and device chip id (CID) are 
public information.
Moreover, legitimacy information items in this category are sometimes not uniquely bound with a device but shared by multiple devices.
For example, in the Alink platform, the legitimacy information is a tuple which consists of two ``confidential'' numerical strings, namely \texttt{Key} and \texttt{Sign}.
We found that obtaining the tuple is
extremely easy -- a bunch of such credentials are available in
the official GitHub repositories of both the Ali company\footnote{\url{https://github.com/alibaba/AliOS-Things}}
and the cooperative manufacturers\footnote{\url{https://github.com/espressif/esp8266-alink-v1.0}}.

\label{sec:mac}
\paragraph{Guessable information} is the information which can be guessed by brute-force.
MAC addresses are a typical kind of guessable information, because the first three bytes in a MAC address are
usually fixed for a manufacturer~\cite{MAC}.
Moreover, manufacturers often allocate a block of consecutive MAC addresses
to the products of the same device model.
Thus,
there remains only two or three bytes for the attacker
to brute-force.
We detail an experiment on Alink devices in Section~\ref{sec:mactest}, in which
we successfully guessed more than 7,181 valid MAC addresses.
Moreover, 
if the adversary can be in the WiFi-range of a victim device,
he can simply eavesdrop the MAC address of the device by sniffing wireless probe requests~\cite{wifisniffer}. Note that this is a fundamental drawback of the WiFi protocol.



\paragraph{Hard-coded information} is unpredictable, immutable, and inherent to a device.
For example, a long \id embedded in the device hardware
is a typical kind of hard-coded information for \hid platforms.
In addition, for some \cid platforms, hard-coded information (e.g., a serial number (SN)) is also
incorporated in the generation of a device ID. 
Although hard-coded information cannot be obtained easily,
it is immutable.
Once this information is leaked, the victim device becomes vulnerable forever.

To get this information, the adversary needs local access to the victim device.  
For example, the adversary can get
the device's hard-coded device ID by sniffing the device-app traffic in the device's LAN during the device discovery phase (Section~\ref{sec:deviceid}). 
In case the adversary is using the device on behalf of the home owner, he can find the hard-coded device ID in a log maintained by the mobile app.  


\label{sec:scene}
We now discuss the feasibility of physically accessing a victim device and the adversary's
incentive to employ the discovered exploits.  
First, the ownership of a device
can be changed if the device gets resold or decommissioned~\cite{blockchainIoT}.
For example, increasingly popular smart home manufacturers such as Samsung and Apple
provide certified refurbished devices on the on-line outlet stores or through Amazon.
The previous owner can easily extract the hard-coded credentials before re-selling the  device.
We have successfully obtained the \id of a Samsung POWERbot R7040 Robot Vacuum in our experiments.
If we resell or return this device, we can control this device remotely
even after it is sold to another user.

Second, a recent study shows that 60\% of guests would actually pay more
for a vacation rental home with smart home features~\cite{airbnbservey}.
Thus, vacation rentals and hospitality service providers like Airbnb.com and Ziroom.com have been collaborating with smart home providers to   
equip an increasing number of smart home devices (e.g., smart locks, cameras and TVs) in their apartments and hotels~\cite{airbnblock}. 
For instance, JD has worked with Ziroom to deploy Joylink smart home devices
in Ziroom rental rooms~\cite{Rental}.
If the adversary ``legitimately'' rents a vacation home for one night and extracts the \id
of the home's smart lock, he could remotely open the lock at will in the future. 
This poses a serious threat to the safety of other tenants. 
We have successfully conducted such a remote hijacking attack against an Alink device (i.e., a KAADAS smart lock with model KDSLOCK001) in lab environment.

\vspace*{-2mm}
\section{Analysis Methodology}
\vspace*{-1.5mm}

The discovered exploits leverage a set of design flaws in the interactions among the three entities.
This section elaborates the vulnerability analysis methodology we used to identify these design flaws. 


\vspace*{-3mm}
\subsection{Deciphering Communication}
\vspace*{-2mm}
\label{sec:reverse}
To protect user privacy,
smart home platforms usually encrypt the communication among
IoT devices, mobile apps, and IoT clouds.
We must decrypt the communication traffic before we can study the interactions.
This imposes significant challenge for us because some platforms are
close-sourced (\eg~~XiaoMi and TP-LINK).

\paragraph{Cloud-App Communication.}
A simple network sniffer confirms that most platforms adopt TLS, 
and mobile apps are required to verify the validity of the cloud (server) certificate.
The MITM attack on the mobile app side is an obvious choice in deciphering the communication.
However, to launch a MITM attack, we must replace the cloud certificate with one controlled by us.
After analyzing a number of mobile apps, we found that a mobile app usually hard-codes the cloud certificate in its APK file 
without relying on the trust store provided by Android~\cite{pin} (a.k.a. certificate pinning).
If we replace the hard-coded certificate in an APK,
the corresponding app would fail to run due to integrity checking. 
We have addressed this problem by rooting our test smartphone and installing an Xposed module\footnote{\url{https://github.com/Fuzion24/JustTrustMe}}.
This Xposed module is able to hook the certificate checking function so that
we can dynamically manipulate the certificate without compromising the app integrity.

\paragraph{Device-App Communication.}
We have analyzed  the APK files of a number of mobile apps and found that some platforms such as Joylink and MIJIA use a symmetric encryption algorithm to protect device-app communication.
Thus, we can easily extract the communication keys by analyzing the APK files.
Other platforms such as SmartThings adopt TLS for device-app communication. To deal with TLS, 
we have used the same method as used for deciphering cloud-app communication.

\paragraph{Device-Cloud Communication.}
Again, device-cloud communication is protected by TLS.
However, we cannot easily replace the cloud certificate embedded in the firmware on a device 
to launch a MITM attack as is done in mobile apps. 
We had to perform static analysis on the device firmware. 
In particular, we have physically dumped the firmware images of the target devices\footnote{\url{https://www.youtube.com/watch?v=KlV3_HaBpbs}}.
We have manually followed the data and control flows of the cryptographic functions,
and were able to locate the hard-coded certificates in the firmware images.  
We then replaced the hard-coded certificates with  a set of certificates forged by us.  
However, we found that the devices enforce firmware integrity verification
which denies executing any manipulated firmware.
Fortunately, simple reverse engineering confirms that most devices only use the simple cyclic redundancy check (CRC) algorithm to check integrity. 
Therefore, we updated the CRC values to match the manipulated firmware and
successfully booted the firmware images with the forged  certificates. 
As a result, we were able to launch the MITM attack
to decrypt the communication.

\begin{lstlisting}[language={json},caption={JSON Representation of Alink Device Registration Message}, keywords={system, alink,
	jsonrpc, lang, sign, key, time, request, cid, uuid, method, params, model, mac, version, id},label=lst:json]
{"system": {
	"alink": "1.0", "jsonrpc": "2.0", "lang": "en",
	"sign": "3a07945eb6f453e6c0a4032c1184cc87",
	"key": "5gPFl8G4GyFZ1fPWk20m", "time": ""
 },
 "request": {
	"cid": "000000000000000010671484", "uuid": ""
 },
 "method": "registerDevice",
 "params": {
	"model": "JIKONG_LIVING_OUTLET_00003",
	"mac": "60:01:94:A2:D5:7C","
	"version": "0.0.0;APP2.0;OTA1.0"
 },
 "id": 100
}
\end{lstlisting}
\vspace*{-1.5mm}

\begin{lstlisting}[language={json}, caption={JSON Representation of Cloud-Side Response to Alink Device Registration Message}, keywords={result,
	code, msg, data, uuid, id}, label=lst:reply]
{"result": {
	"code": 1000, "msg": "success",
	"data": {
		"uuid": "D66FCB11A731CA2683A6C0DED6CD326D"
	}
 },
 "id": 100
}
\end{lstlisting}

\vspace*{-3mm} 
\subsection{Understanding the Interacting Messages}
Using the aforementioned approaches, we were able to
reveal plain-text network traffic among IoT devices, mobile apps, and clouds.
This greatly simplified our analysis.
Although different platforms adopt different communication protocols,
it is a common practice that messages are encoded using the JSON (JavaScript Object Notation) format,
which is quite self-explanatory. 
For instance, we show a message sent from a device to
the cloud of the Alink platform in Listing~\ref{lst:json}.
As indicated by the \emph{method} field,
this message is used to register the device.
The device legitimacy information being sent includes \texttt{Sign} and \texttt{Key}.
The device identity information being sent includes \texttt{CID}, \texttt{model}, and \texttt{MAC}.
The respond message is shown in Listing~\ref{lst:reply}.
As we can see, the \id is returned in the \texttt{uuid} field.

\vspace*{-1mm}
\subsection{Phantom Devices}
\vspace*{-1.5mm}

\label{sec:analysis}
\label{sec:phadevice}


We investigated the interactions from three aspects.
First, we tested whether each entity strictly maintains its state machine,
which means an entity should only accept interacting requests acceptable in its current state.
For example, as shown in Figure~\ref{fig:s1}, when an IoT cloud is working in state 4, 
it should deny the request from a device to bind itself to another user account.  
Second, we tested whether the three entities always
stay in a legitimate 3-tuple state combination 
(see Appendix~\ref{sec:appendix1}).
Third, we adjusted the parameters, especially those used in authentication,
of the normal interacting requests and observed the responses.
Our goals is to discover whether the receiving entity
of each request conducts proper authorization checking.


\label{sec:phantom}
To make this happen, we need to be able to craft JSON messages
and send them to the receiving entities.
However, we cannot arbitrarily change the requests of a real device.
To cross this barrier, we created and ran a phantom device (program) that mimics a real device to assist our analysis.
A \pha device is constructed as follows.
Some smart home platform providers like Samsung, JD and Ali open-source their device-side SDKs and demo programs,
which include the same communication logic as a real product.
We simply reused them to build our \pha devices.
On the other hand,
XiaoMi and TP-LINK use proprietary SDKs, and 
we had to reverse-engineer the firmware we obtained from real devices,
and implement programs to imitate the original communication functions.


With the help of \pha devices,
we could arbitrarily adjust the parameters of request messages.
In this way, we could trigger 
unexpected state transitions and manipulate/remove the authentication fields of a request
to perform black-box testing against an IoT cloud.
We shortly report our findings in Section~\ref{sec:flaws}. 

\label{sec:idgen}
The \pha devices not only facilitated interaction analysis, but also
helped us figure out the relevant internal logic of an IoT cloud, which was completely opaque to us.
For instance, we used a \pha device to test and confirm which device identity 
information is used in the generation of a \id.
Specifically, for the Alink platform, 
we changed the value of each field appeared in Listing~\ref{lst:json} and
recorded the corresponding returned \ids from cloud.
Finally, we compared the received \ids
to infer which fields influence the generation of a \id.
We concluded that the fields \texttt{model}, \texttt{MAC} and
\texttt{CID} uniquely determine a \id.
In other words, there is a one-to-one mapping between the (\texttt{model}, \texttt{MAC}, and \texttt{CID}) tuples and device IDs in the Alink platform.
We used a similar method to test all the studied platforms and the results are summarized 
in Table~\ref{tab:info}.

\vspace*{-2mm}
\section{Identified Design Flaws}
\label{sec:flaws}
\vspace*{-1.5mm}  

\begin{figure*}[t]
\centering
\includegraphics[width=0.184\textwidth]{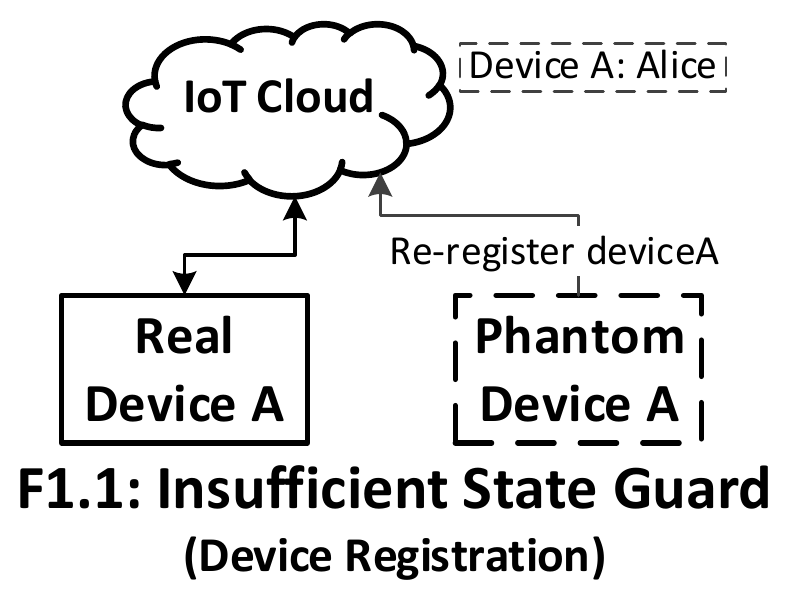}%
~
\includegraphics[width=0.264\textwidth]{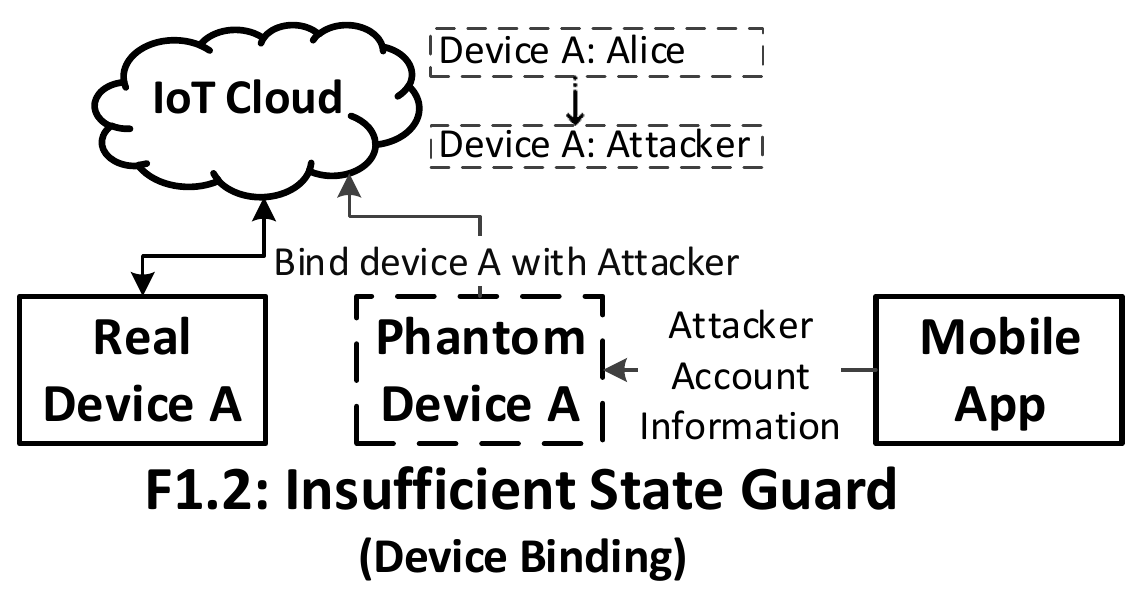}%
~
\includegraphics[width=0.184\textwidth]{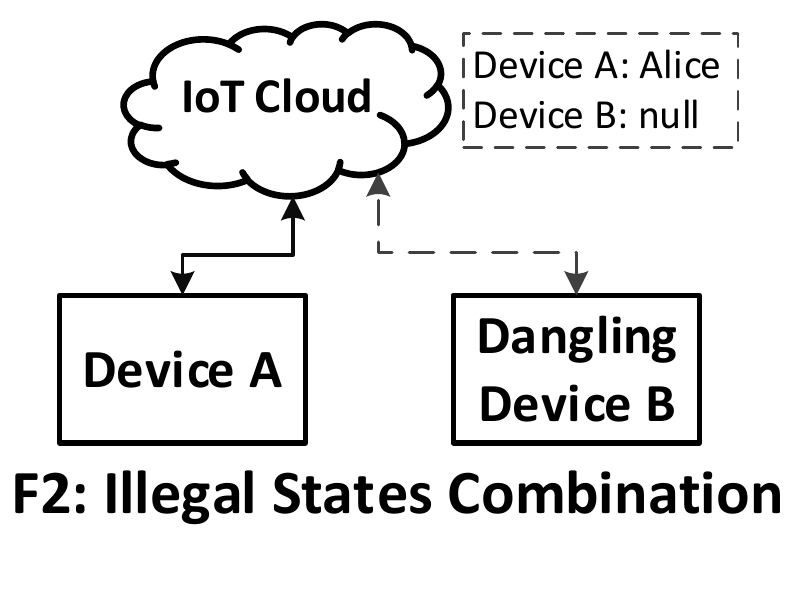}%
~
\includegraphics[width=0.183\textwidth]{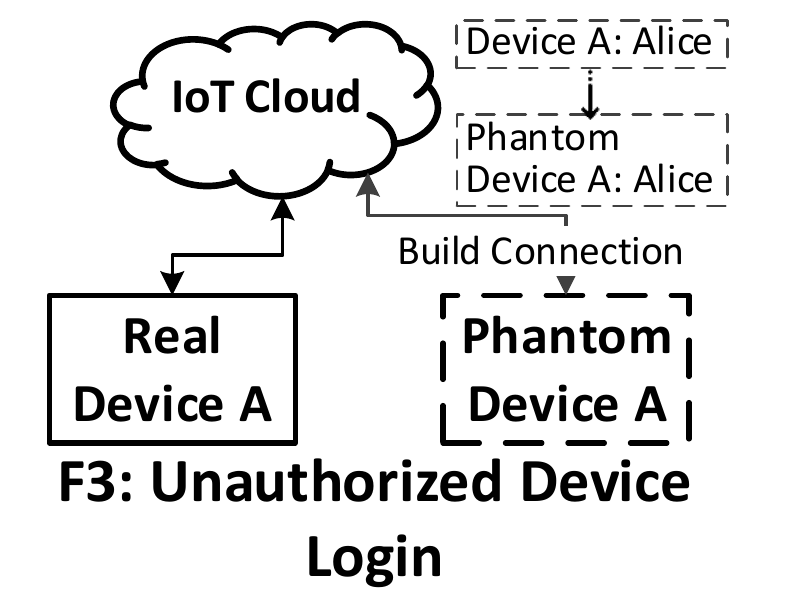}%
~
\includegraphics[width=0.183\textwidth]{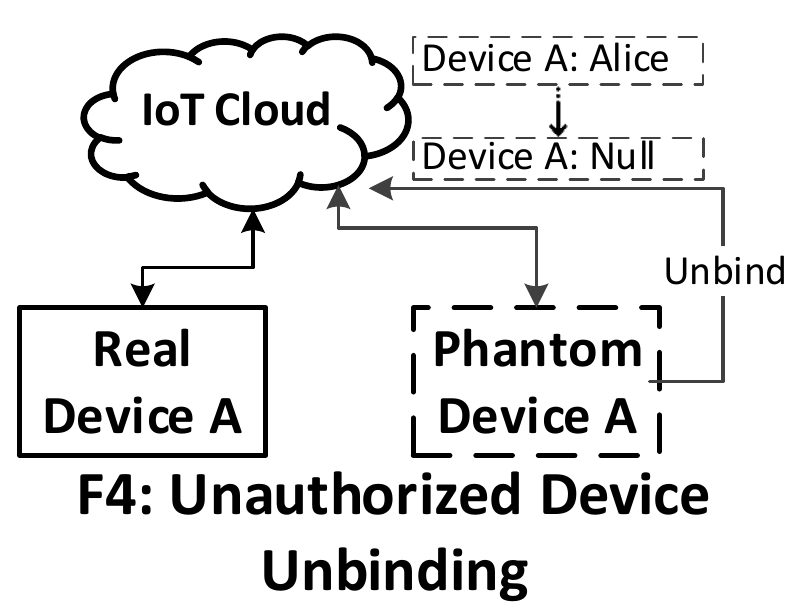}%
\caption{Identified Design Flaws}
\vspace*{-3mm}
\end{figure*}


Using the analysis methodology presented in Section 4, 
we have discovered four kinds of design flaws,
and we have shared them with the providers of the five platforms we investigate. 
These design flaws are summarized in Table~\ref{tab:flaw}.

\begin{table}[t]
  \caption{Device Identity/Legitimacy Information}
  \vspace*{-1mm}
  \begin{adjustbox}{max width=\columnwidth}
  \begin{tabular}{c|c|c|cc}
    \toprule 
     &\textbf{Platform}&
     \textbf{Identified Flaws}&
     \textbf{Exploited Flaws}&
     \textbf{Applicable Attacks}\\
    \midrule
    \multirow{5}{*}{\tabincell{c}{\\[-3em]\textbf{\cid }\\\textbf{Platform}}}&Alink&\tabincell{c}{F1.1, F1.3\\F2, F3, F4}&\tabincell{c}{F1.1, F1.3, F2, F3, F4\\F1.1, F1.3, F3\\F1.1, F1.3, F3, F4\\F1.1}&\tabincell{c}{Remote Device Hijacking\\Remote Device Substitution\\Remote Device DoS\\Illegal Device Occupation}\\
    \cline{2-5}
    &Joylink$^*$&\tabincell{c}{F1.1, F1.3\\F2, F3}&\tabincell{c}{F1.1, F1.3 F3\\F1.1}&\tabincell{c}{Remote Device Substitution\\Illegal Device Occupation}
    \cr\hline\hline
    \multirow{6}{*}{\tabincell{c}{\\[-2em]\textbf{\hid }\\\textbf{Platform}}}
    &KASA&F1.2, F1.3, F3&\tabincell{c}{F1.2, F3\\F1.3, F3\\F1.2}&\tabincell{c}{Remote Device Hijacking\\Remote Device Substitution\\Remote Device DoS}\\
    \cline{2-5}
    &MIJIA&F1.2, F1.3, F3&\tabincell{c}{F1.2, F3\\F1.3, F3\\F1.2}&\tabincell{c}{Remote Device Hijacking\\Remote Device Substitution\\Remote Device DoS}\\
    \cline{2-5}
    &SmartThings$^\dag$&F1.2, F1.3&F1.2&Remote Device DoS\\
    \bottomrule
  \end{tabular}
  \end{adjustbox}
\flushleft
\scriptsize{
*: Joylink platform does not support device-side unbinding request.\\ \dag: SmartThings cloud performs authorization checking on device login request.}
\label{tab:flaw}
\vspace*{-3mm}
\end{table}

\paragraph{F1: Insufficient State Guard.}
\label{sec:F1}
We found that none of the three entities correctly guard their state machines.
This could lead to severe consequences. 
Since IoT clouds are responsible for security-critical services such as device identify management, IoT clouds can be most affected. 
In the state machine of an IoT cloud (Figure~\ref{fig:s1}),
when the cloud is working in state 4 (running),
ideally it should only accept status upload requests (edge 6) or
device unbinding requests (edge 3). 
Unfortunately, we found that the IoT cloud also accepts other requests. 
Depending on which request is accepted incorrectly when
the IoT cloud is in state 4,
we break down flaw F1 into three sub-flaws. 


{\bf F1.1:} 
This flaw is specific to \cid platforms.
An attacker, having all the device identify information,
can use a \pha device to send a registration request to the cloud,
which is fooled to return the corresponding \id to the attacker (Figure 3F1.1).

{\bf F1.2:} 
This flaw is specific to \hid platforms.
An attacker can use a \pha device to send a binding request that 
links the device (identified by \id) with the attacker's account (Figure 3F1.2).
Note that in \hid platforms, the binding request is sent from the device
and the cloud unconditionally accepts the binding request (see Section~\ref{sec:interactionprocess}).
As a result, the \pha device can bind the attacker's account to the victim device.

{\bf F1.3:} 
The IoT cloud accepts device login requests even if
it is in state 4.
This flaw is a pre-condition of flaw F3 which
we will describe shortly. 




\paragraph{F2: Illegal State Combination.}
\label{sec:F2}
We found that the three entities sometimes stay in unexpected and illegal state combinations.
One root cause is flawed synchronization among them.
When an illegal state combination is exploited, security can be violated.
For instance, ideally,
when a user retires a smart home device,
he should reset and unbind the device,
making all of the three entities go back to their initial states (\ie~state combination (S1, S1, S1)).
However, for \cid platforms, if the user unbinds the device without firstly resetting it,
a connection with the cloud
is still retained and the state combination actually switches to (S1, S4, S1), which is illegal based on our table in Appendix~\ref{sec:appendix1}. 
Since the device is in this illegal state combination,  
we call it a dangling device (Figure 3F2). 
Now, since the IoT cloud is in state 1, 
if the attacker remotely issues a request to
register this device,
the request is allowed and the cloud transfers to state 2.
For the same reason, if the attacker continues to send
a request to bind the device, the cloud accepts the request
and transfers to state 4 (state 3 is skipped because
a connection is still maintained with the victim device).
At this time,  if the attacker sends a control 
command to the device,
the cloud will mistakenly forward the command to the retired  device.
This essentially causes a device hijacking attack.


\paragraph{F3: Unauthorized Device Login.}
\label{sec:F3}
A connection is maintained between the device and the IoT cloud after
device login.
Ideally, the cloud should only allow a login request if the request is issued
from the device that is bound with the owner account.
However, we found that the IoT cloud does not perform any account-based authorization check
during device login. 
In other words, the connection is decoupled from the user account. 
Consequently, 
when the attacker uses a \pha device to login with the \id of the victim device,
the cloud is fooled into establishing a connection with the \pha device (Figure 3F3).
As a necessary condition of this flaw, the cloud must accept
device login requests in state 4, which is exactly what Flaw f1.3 states.


\paragraph{F4: Unauthorized Device Unbinding.}
\label{sec:F4}
Ideally, only the user who holds an account currently bound to a device 
has the privilege to unbind the device.
This is true if unbinding operations are conducted on mobile apps,
which indeed include user credentials in unbinding messages.
Unfortunately, for \cid platforms, device unbinding can also be achieved
on the device side. Based on our analysis, device-side unbinding commands do not included any user credentials.
As a consequence, an attacker can build  a \pha device to forge an unbinding request
using device-side API. The binding relationship between victim user's account
and the device is then revoked without the user's awareness (Figure 3F4).

\vspace*{-4mm}
\section{Flaw Exploitation}
\vspace*{-3mm}

Exploiting various combinations of the identified design flaws, an attacker can
launch a spectrum of attacks,
including remote device substitution, remote device hijacking, remote device DoS, illegal device occupation, and firmware theft. 
In the following, we first describe the experimental setup.
Then we detail two most severe attacks revealed in this paper -- how to remotely substitute and hijack a victim device, respectively.  
We also discuss the other attacks. 
In Table~\ref{tab:flaw}, we summarize the 
set of particular attacks, as well as the design flaws exploited by each attack.   
To visually demonstrate some of the discovered exploits, 
we also recorded two videos\footnote{\url{https://youtu.be/MayExk_PKhs}}\footnote{\url{https://youtu.be/fufEAtQq2_g}}.


\vspace*{-3mm}
\subsection{Experimental Setup}
\vspace*{-1.5mm}
\label{sec:setup}

All the PoC (Proof of Concept) attacks were conducted within lab environment without influencing legitimate users. 
The tested devices include smart plugs, IP cameras, WiFi bulbs, cleaning robots and smart gateways, covering all the studied platforms. 
These devices are shown in Figure~\ref{fig:experiments} and we also
list them in Appendix~\ref{sec:appendix4}.

\begin{figure}[t]
\begin{center}
\includegraphics[width=0.8\columnwidth]{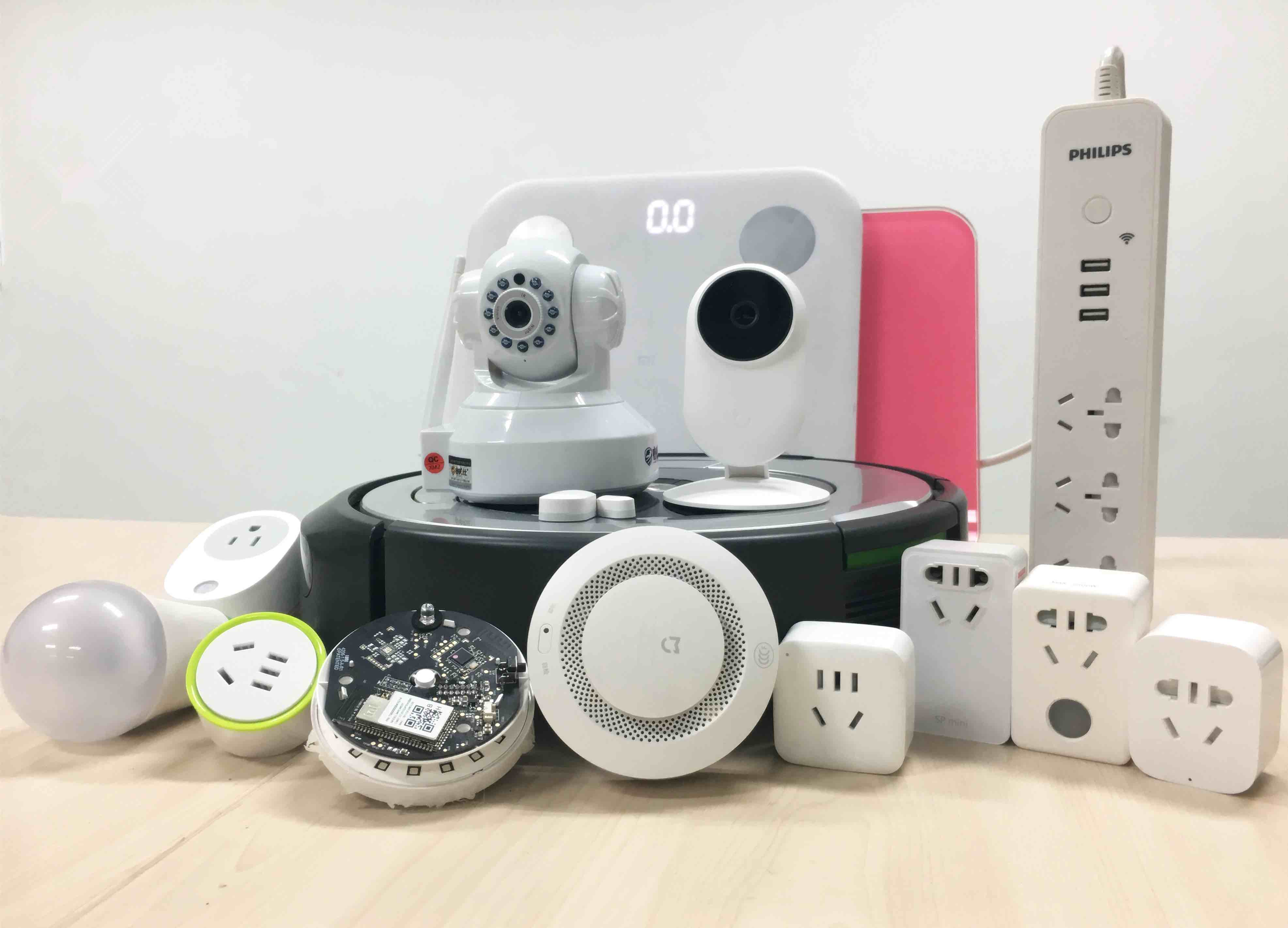}
\caption{Smart Home Devices Used in Our Experiments}
\label{fig:experiments}
\vspace*{-9mm}
\end{center}
\end{figure}

\paragraph{Obtaining Device Identity and Legitimate Information.}
We need device identity and legitimacy information
listed in Table~\ref{tab:info} to forge a \emph{phantom} device.
As mentioned earlier, the difficulty of obtaining an information item differs.
\label{sec:mactest}
In the following, we use an Alink device (Philips smart plug with model SPS9011A) and a TP-LINK device (WiFi Bulb with model LB110)
to represent \cid and \hid devices, respectively.
We describe how to obtain their identity and legitimacy information. 
For other platforms,
similar approaches can be adopted.

For the Alink device, we collected its legitimacy information (\texttt{Key} and \texttt{Sign})
very easily through the public repositories mentioned in Section~\ref{sec:fea}.
The Alink device uses \texttt{model}, \texttt{MAC}, and \texttt{CID} as its identity information.
For \texttt{model} and \texttt{CID}, which are fixed for a
specific device type,
we extracted them from a log maintained by the corresponding mobile app. 
For the MAC address, we obtained it by brute-force attack.
Specifically,
we bought two test devices (Philips smart-plugs) and
recorded the MAC address of the first one (``3C:2C:94:0B:56:69'') manually.
The first device served as the attacker's device in our experimental setting.
Then we tried to guess the MAC address of the other device, which played the ``victim device'' role in our experiments.
After 21,692 mutations to ``3C:2C:94:0B:56:69'', we ``hit'' the victim device's MAC
address, which turned out to be ``3C:2C:94:0B:AB:25''.

To further demonstrate that collecting MAC addresses of real devices is not very difficult, 
we mutated the lower two bytes of ``3C:2C:94:0B:56:69'' to get
65,536 different MAC addresses. 
Then we used these MAC addresses with other
fixed identity/legitimacy information to register the victim device.
Every registration request returned us a \id.
We further wrote a probing script
running on the mobile to automatically bind the returned \ids to our test user account.
If a \id has already been bound to an account,
the cloud will refuse the corresponding binding request with an error message. 
We found that 7,181 \ids had already been taken by real users.
It means the corresponding 7,181 MAC addresses were actively being used by real devices and
they could potentially become victims in our attacks.

\emph{Ethical Consideration.} 
In order to prevent potential influence on legitimate users,
before we ran the script that tests active MAC addresses, 
we first evaluated the impact on a test device in 
lab environment. 
Specifically, we first simulated a legitimate user by normally binding a tested device to an account. Then we used a \pha device with the same identity and legitimacy information as the test  device and got the same \id from the cloud.
We then used another account to try to bind the device as was done by the probing script.
During this process, we found that 
the test device could still be operated normally and was not unbound.

For the TP-LINK device, since its identity information (i.e., \id) and 
legitimacy information (i.e., \texttt{hwid} and MAC address) are hard-coded,
we had to physically extract them.
Specifically, 
we collected \ids, \texttt{hwid} and MAC address by launching
a MITM attack that intercepts the device-app communication.


\paragraph{Building Phantom Devices.}
With all the required information available, we implemented \pha devices
using Python.
As discussed in Section~\ref{sec:phadevice}, when the device-side SDK is
available (e.g., on platforms provided by Samsung, JD and Ali),  we directly incorporated them into our program.
Otherwise, our Python program mimics the behavior of a device.
The behavior knowledge is obtained by reverse-engineering the firmware extracted from real devices (XiaoMi and TP-LINK).
In total, we implemented five kinds of \pha devices for
Samsung, Joylink, Alink, XiaoMi and TP-LINK, respectively.
They were implemented by 22, 17, 14, 61 and 72 
lines of code (excluding SDK functions if any), respectively.

\paragraph{Network Configuration.}
We placed the target devices and the \pha devices in two separate LANs behind NAT-enabled routers.
As a result, the target devices and the \pha devices cannot
communicate with each other directly. 
This setting resembles real-world scenarios where a 
remote attacker 
does not have access to the LAN of the victim.  

\vspace*{-2mm}
\subsection{Remote Device Substitution}
\label{sec:subattack}
\vspace*{-1.5mm}

In this subsection, we showcase how an attacker can remotely replace the victim's device with a \pha device under his control.
To simplify the presentation, in the following, we use $Alice$ to denote the victim/legitimate user and $Trudy$ to denote the attacker.
We use sequence diagrams to represent
the interactions among the three entities.
An attack happens when $Trudy$ interferes with the normal interaction diagrams.
In Appendix~\ref{sec:appendix3}, we show the normal interaction diagrams for all the studied platforms.

\begin{figure}[t]
\includegraphics[width=\columnwidth]{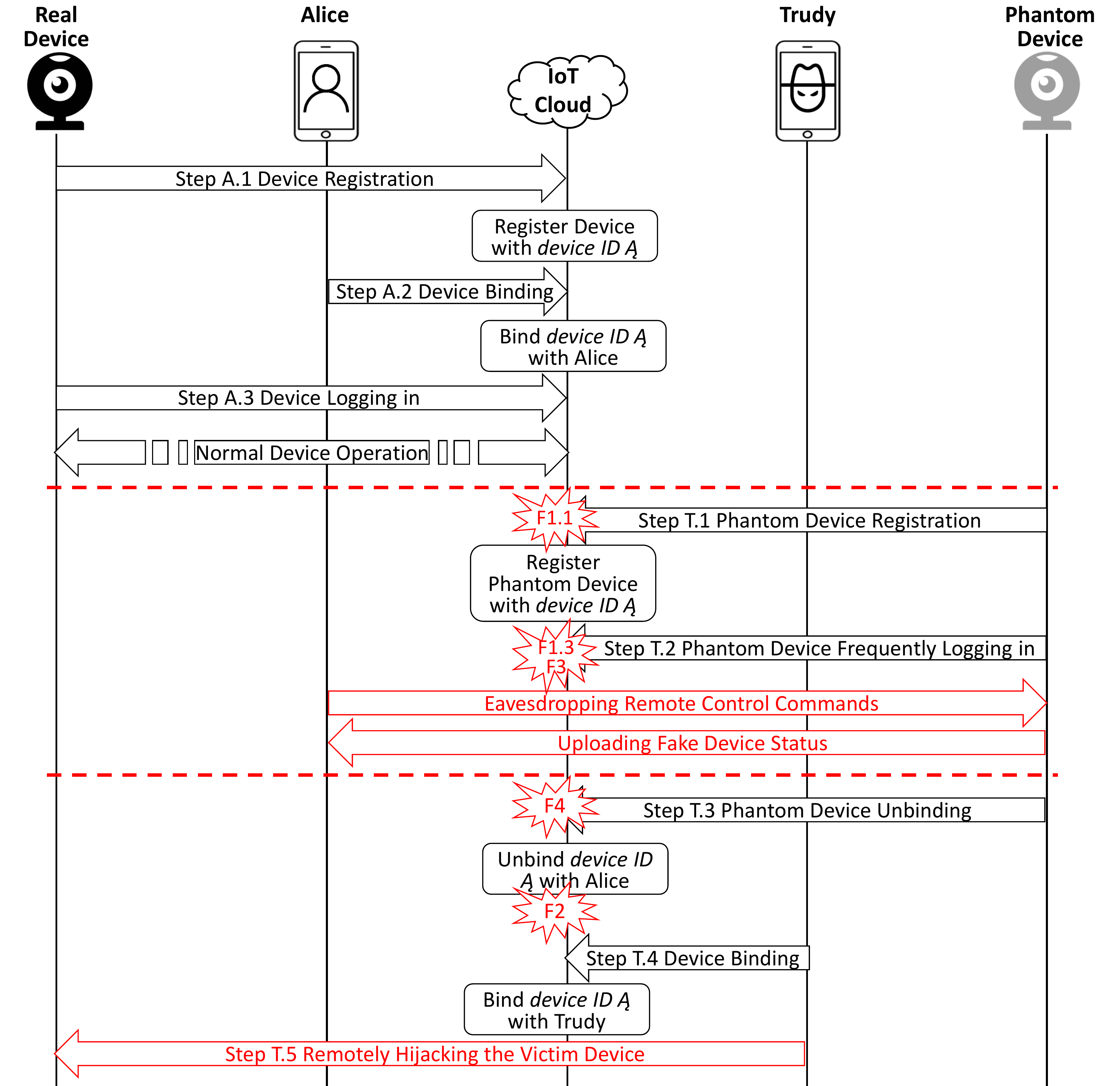}
\caption{Remote Attacks on \cid Platforms}
\label{fig:Cloud_id_attack}
\vspace*{-5mm}
\end{figure}

\paragraph{Attack Workflow (\cid).}
On the top of Figure~\ref{fig:Cloud_id_attack} (above the highest dashed red line),
we show the normal workflow of how $Alice$ uses her IoT device 
on a Type I platform. 
After $Alice$ provisions the device with a WiFi credential,
the device sends its legitimacy credential and device identity information
to the cloud to get registered (Step A.1).
Based on the device identity information, the cloud registers the device
with a \id $\mathcal{A}$
and binds it to $Alice$'s account (Step A.2).
After the device logs in (Step A.3), $Alice$ can control the device with her account.

Then, the attacker, $Trudy$, kicks in as shown in the middle of the figure (between the two dashed red lines).
She first lets the phantom device send the same device registration request as used in step A.2 to the cloud (Step T.1).
Due to F1.1, the cloud accepts this request and
registers the \pha device with the same \id $\mathcal{A}$,
but still keeps \id $\mathcal{A}$ bound with $Alice$.
At this moment, $Alice$ actually binds the \pha device and real device at the same time.
Then $Trudy$ could leverage the flaw F1.3 and F3 to log in a \pha device without $Alice$'s account information (Step T.2).
Since the \pha device has the same \id as the real device,
the cloud disconnects the original connection with the real device
and establishes a new connection with the \pha device.
However, when the real device does not receive the heartbeat message for a while,
it automatically logs into the cloud again and puts the \pha device offline.
Now, the real device and the \pha device are in fact competing for connection with the cloud.
To win the competition, $Trudy$ configures the \pha device to login very frequently.
As a result, the \pha device always wins.
$Alice$ now still appears to ``control'' a device through
her mobile app,
although this device has actually been replaced by the \pha device under the control of $Trudy$.

\begin{figure}[t]
\includegraphics[width=\columnwidth]{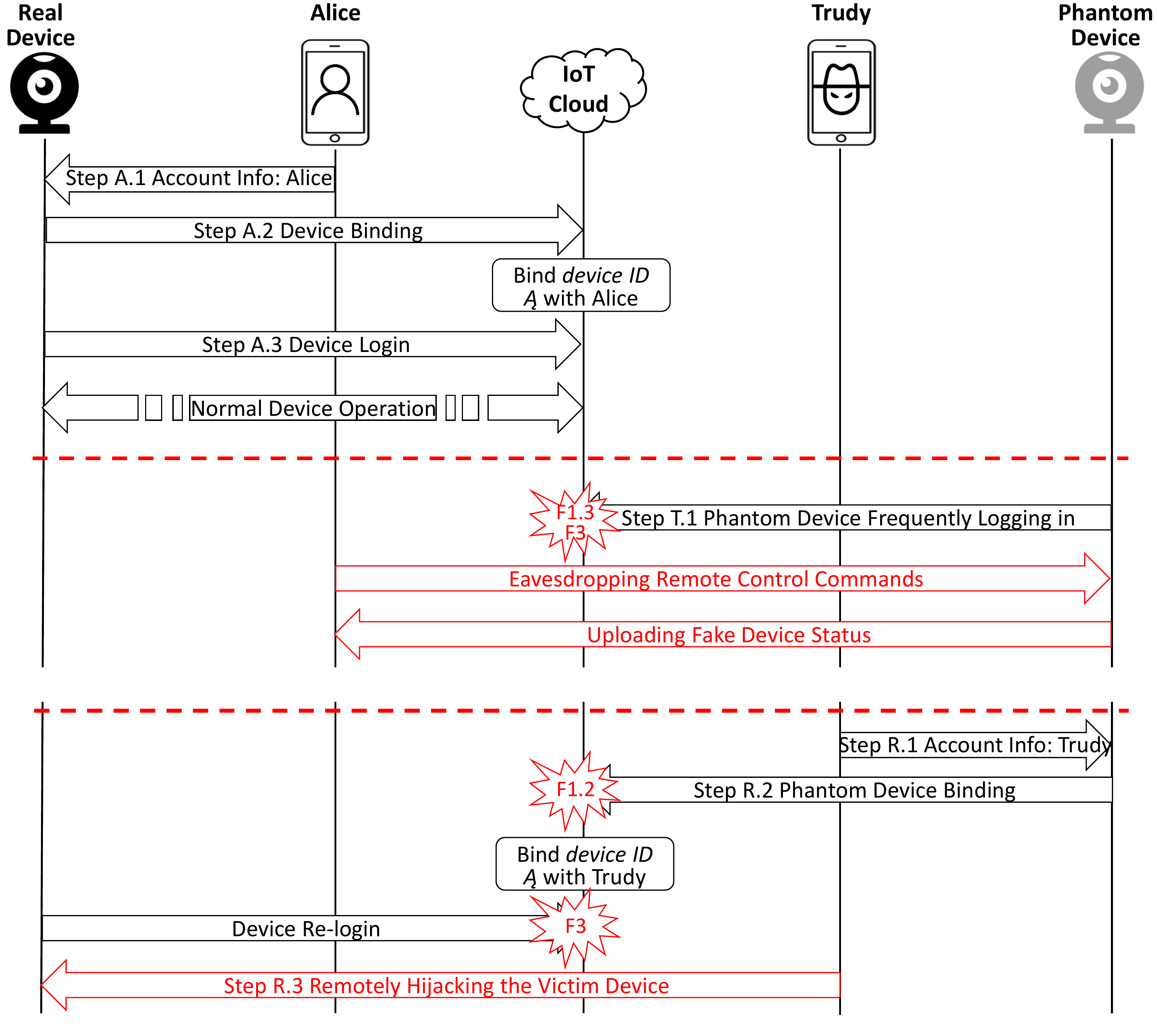}
\caption{Remote Attacks on \hid Platforms}
\label{fig:Hard_code_attack}
\vspace*{-5mm}
\end{figure}

\paragraph{Attack Workflow (\hid).}
Similarly, the top part of Figure~\ref{fig:Hard_code_attack} (above the highest dashed red line) shows
the normal workflow of how $Alice$ uses her device on 
a \hid platform. 
After her mobile app sends her account information to the device (Step A.1),
the device sends the binding request with
\id and legitimacy information, as well as the account information to the cloud (Step A.2).
The cloud binds the \id $\mathcal{A}$ to $Alice$'s account. 
After the device logs in the cloud (Step A.3), $Alice$ can control/monitor the device with her mobile app.


In the middle of the figure (between the two dashed red lines), 
$Trudy$
launches the remote device substitution attack.
Enabled by flaws F1.3 and F3, 
she lets the phantom device successfully log into the cloud with the same \id (Step T.1). 
At this time, the \id $\mathcal{A}$ is still bound to $Alice$'s account.
Like in the \cid platform,
the \pha device maintains a connection with the cloud by periodically logging in.
In this way, the attacker secretly substitutes $Alice$'s device with
a \pha device under her own control.


%

\paragraph{Attack Consequence: Privacy Breaches.} 
In normal operations,
when $Alice$ uses her mobile app to send a remote control command to the cloud,
the cloud forwards the command to the ``device''.
Unfortunately, in the remote substitution attack,
the real device has been replaced by a \pha device controlled by the attacker.
As a result, all the control commands from $Alice$
are exposed to the \pha device and further to $Trudy$,
leading to a privacy breach. 
For example, if $Trudy$ substitutes a smart plug, he could
know when $Alice$ turns on/off the smart plug.
This information could be used to infer whether 
$Alice$ is at home. 

\paragraph{Attack Consequence: Falsified Data.} 
In normal operations, the real device updates its sensor readings to the cloud and the result 
is reflected in $Alice$'s mobile app. 
Unfortunately, in the remote substitution attack, the sensor readings are
sent from the \pha device.
This gives $Trudy$ an opportunity to manipulate the sensor readings sent 
to $Alice$, thus deceiving or misleading $Alice$.
For example, we tested a XiaoMi smoke alarm (model: Fire Alarm Detector) and a Alink smart lock (model: KAADAS KDSLOCK001).
If the smoke alarm detects a thick smoke in the room, 
the smart lock will be unlocked automatically to open the window/door.
We used remote device substitution attack to manipulate the sensor readings
of smoke alarm and successfully unlocked the smart lock.
This leads to serious consequence because $Trudy$ can enter $Alice$'s room at will.


Our attack can also serve as a trigger for the flaws mentioned in
previous works~\cite{celik2018soteria,Kafle2018A}.
Once a less-protected device is substituted by a \pha device and the device is in the chain of a ``routine'', the \pha device can further influence other data sensitive devices.
For example, as mentioned in~\cite{Kafle2018A}, the Nest Cam monitor in a house will automatically switch off when the global ``away/home'' state changes from ``away'' to ``home''. 
$Trudy$ can take advantage of the substitution attack to change this state variable to disable the camera and burglarize the house without being recorded.



\paragraph{Stealthiness Analysis.}
Remote device substitution attack is highly stealthy.
This is because $Alice$ always sees the
device to be online in her smartphone (although it is a \pha device). 
However, if $Trudy$ feeds the \pha device with sensor readings that
excessively deviate from normal,
$Alice$ (if she is security-savvy) might become suspicious of the dramatic change. 




\vspace*{-3mm}
\subsection{Remote Device Hijacking}
\label{sec:hijacking}
\vspace*{-1mm}
$Trudy$ can further remotely control $Alice$'s device by exploiting
more flaws. We call this remote device hijacking attack.

\paragraph{Attack Workflow (\cid).}
Continuing from Step T.2,
device hijacking attack is depicted
at the bottom of Figure~\ref{fig:Cloud_id_attack} (below the lowest dashed red line).
At this moment, the \pha device has already logged in the cloud.
Due to flaw 4, $Trudy$ is able to send a device-side
unbinding request via her \pha device to the cloud (Step T.3), which puts
the real device in dangling status (due to F2).
Finally, $Trudy$ binds the device with her account (Step T.4).
As a consequence, $Alice$'s device is connected with the cloud whereas
$Trudy$ is able to control the device on her smartphone.

\paragraph{Attack Workflow (\hid).}
Unlike \cid platforms,
to carry out a hijacking attack against \hid platforms,
$Trudy$ does not need to continue from the success of a remote substitution attack.
Instead, she starts attacking from scratch, as depicted
at the bottom part of Figure~\ref{fig:Hard_code_attack} (below the lowest dashed red line). 
$Trudy$ starts by using her mobile app to send her account information 
to the \pha device (Step R.1).
Next, the \pha device, which has $Alice$'s device and other legitimacy information,
will send a binding request with $Trudy$'s account information (Step R.2) to the cloud.
Due to F1.2, the cloud is fooled into accepting the binding request.
Now, the ownership of the device has changed to $Trudy$ from the view point of the cloud. 
The cloud then terminates the connection from the real device.
However, the real device has re-connecting mechanism that
continuously restores lost connection to the cloud.
Due to F3, the cloud does not verify whether the account information ($Alice$)
matches the current device owner ($Trudy$) or not.
Therefore, the re-connecting request from the real device 
can be successful.
At this point, $Trudy$ could completely control $Alice$'s device (Step R.3).



\paragraph{Attack Consequence.}
The remote device hijacking attack allows $Trudy$
to bind her account to $Alice$'s device.
As a result, she can harvest the sensor readings in $Alice$'s home.
She can also send remote commands to control $Alice$'s device.
In our experiment,
we successfully hijacked a Alink IP camera (model: RIWYTH RW-821S-ALY).
As a result, we can view the victim's IP video feeds secretly,
greatly threatening victim's privacy.




\paragraph{Stealthiness Analysis.}
Since $Alice$'s device has been hijacked,
she can no longer talk to her device. 
It could raise an alert for $Alice$ if she is security-savvy.
Average Joe may simply regard it as a service failure and
rebind his user account.
It is worth mentioning that even for security-savvy users,
it is not easy to trace back to the attacker.
Only the IoT cloud has some clues to trace back to the attack origin.

\vspace*{-3mm}
\subsection{Other Security Hazards}
\vspace*{-1mm}
\label{sec:others}

\vspace*{-1mm} 
\subsubsection{Remote Device DoS}
\vspace*{-1.5mm}
As a basic security measure, IoT clouds only allows authorized users to control a device.
If an attacker can unbind a target device from its legitimate user, the 
target device cannot be operated anymore,
essentially leading to device denial of service (DoS) attack.
To launch this attack, the attacker does not need to exploit many flaws.
In particular,
for \cid platforms, after the attacker sends the device-side unbinding command (Step T.3), as shown in Figure~\ref{fig:Cloud_id_attack}, the cloud directly revokes the binding relationship between the victim user and the device.
For \hid platforms, as shown in  Figure~\ref{fig:Hard_code_attack},
after the attacker leverages flaw F1.2 to bind a \pha device with his account (Step R.2),
the target device is unbound.


Note that since remote device DoS attacks require less flaws, the attack is applicable to more platforms.
For instance, the Samsung SmartThings platform, which is immune to 
remote device substitution/hijacking attacks, is vulnerable to  remote device DoS attacks. 
This is because the SmartThings platform is not subject to F3 which is essential for 
remote device substitution/hijacking attacks.
However, F3 is not required in remote device DoS attacks.
We will explain why SmartThings is not subject to F3 in Section~\ref{sec:SmartThingsF3}.

\vspace*{-2.5mm}
\subsubsection{Illegal Device Occupation}
\label{sec:DO}
\vspace*{-1.5mm}
Although a device may be shared with multiple users, only one user account is allowed to be bound to a smart home device.
If the attacker can predict the \ids of unsold devices and use \pha devices to
bind them with valid user accounts, these devices cannot be bound again after being sold.
We call this attack illegal device occupation.
In essence, this attack makes new devices unavailable to legitimate consumers.
Note that this attack only applies to \cid platforms since attackers
can predict device identity information.
In \hid platforms, long and unpredictable \ids are hard-coded in devices.
Attacks can no longer learn anything about \ids until the device is sold.

\vspace*{-3mm}
\subsubsection{Firmware Theft}
\label{sec:FT}
\vspace*{-2mm}
To protect intellectual property (IP),
most IoT manufacturers employ certain tamper-resistant techniques to protect their products,
including enforcing read-only property on flash chips that store proprietary firmware. 
However, with leaked-out firmware, the attacker can
reverse-engineer it, causing IP theft and 
harming the corresponding manufacturers. 
By exploiting OTA updates available on most IoT devices, 
our firmware theft attack is able to bypass the aforementioned protections.
By forging different kinds of \pha devices,  
the attacker can issue OTA update requests in bulk, and thus he can harvest hundreds of firmware images in seconds.

In the experiments, we created phantom devices to ``emulate'' 1,355 kinds of Alink devices, 543 kinds of Joylink devices,
118 kinds of XiaoMi devices,
23 kinds of  SmartThings devices, and 
18 kinds of TP-LINK devices.
Eventually, we were able to collect 63 firmware images
from the Alink platform, 37 from the Joylink platform,
89 from the MIJIA platform, and 16 from the KASA platform.

\vspace*{-2mm}
\section{Discussion}
\vspace*{-2mm}
\label{discussion}

Although we only studied five popular cloud-based smart home platforms,
our findings can be generalized to other 
smart home devices (\eg~hub-connected devices),
and have implications to platforms that are non-cloud-based. 
We also 
discuss the root causes of the identified vulnerabilities,
and suggest several potential defensive approaches to mitigating the discovered exploits. 
Finally, we discuss the impact of the discovered exploits on commercial competitions.

\vspace*{-3mm} 
\subsection{Impact on Hub-Connected Devices}
\label{sec:hub}
In this paper, we focus
on cloud-connected devices.
However, the discovered exploits are
also applicable to hub-connected devices due to two reasons.

First, the attacker can leverage an already exploited hub to
control hub-connected devices.
For example, a XiaoMi smart gateway is a hub for MIJIA products.
After hijacking the gateway, the attacker can further
control all its connected devices.
Note that Samsung SmartThings is not vulnerable to this attack because a SmartThings hub unbinds all the connected devices
when the ownership of the hub is changed, which
is inevitable in this attack.

Second, by forging a \pha hub-connected device,
it is possible to launch firmware theft and illegal device occupation attacks.
However, since the target devices are behind a hub, it is impossible to remotely hijack or substitute them. 

\vspace*{-2mm}  
\subsection{Implications to Cloud-Free Smart Home Platforms}
\vspace*{-2mm}  
\label{sec:non-cloud} 


\paragraph{HomeKit.} 
HomeKit~\cite{HomeKit} is Apple's proprietary smart home platform.
Compatible devices run the HomeKit accessory protocol to directly talk to
mobile apps via WiFi or Bluetooth.
Moreover, using the mobile app, users can access home devices
indirectly through a hub device (Apple TV, HomePod or even iPad).
In this case, the cloud relays commands from the mobile app to the Apple TV or HomePod.
Then the Apple TV or HomePod issue commands to home devices locally.
Note outside the home LAN, there is no direct link
to home devices~\cite{HomeKitlink}.
As a result, our attacks are not applicable to HomeKit devices.

\paragraph{DIY Platforms.} 
DIY smart home platforms such as Home Assistant~\cite{HA} and OpenHAB~\cite{OH} 
are open-source projects that focus on building local home automation.
Due to controllable privacy and low cost (as low as the price of a Raspberry Pi), they are
becoming more and more popular among DIY enthusiasts.
In essence, they build private hubs that interact with different home devices.
To support as many devices as possible, these platforms can be extended with components,
which implement device specific logic~\cite{hacomponents}.
While some devices can work by connecting to the hub locally (\eg~Philips Hue),
others cannot work without relying their own cloud backends.
To this end, Home Assistant classifies the smart devices into two types:
devices that interact with third-party clouds, and devices that respond to 
events that happen within Home Assistant. 
For the former type, the hub serves as a proxy of other third-party clouds. 
For example, 
if a user wants to use a SmartThings device through Home Assistant, 
he first registers and binds the device with the SmartThings cloud. 
Then he needs to install a SmartThings plugin for Home Assistant
to connect the device to the hub~\cite{HAS}. 
The plugin stores the user's SmartThings account token 
and delegates all the device requests to the SmartThings cloud.

For smart home devices relying on their own cloud backends,
Home Assistant is actually transparent to the devices
and thus all the exploits discovered in this paper can be applied to them. 
However, our attacks cannot influence devices that only work locally.


\vspace*{-3.5mm}
\subsection{Root Cause Analysis}
\vspace*{-2mm}

Some of the vulnerabilities revealed in this paper are associated with inherent design flaws of smart home platforms.
Some are themselves caused by design challenges of smart homes.
Therefore, some of security flaws cannot be remedied in a straightforward way.

\paragraph{Ownership Transfer.}
A natural assumption that smart home manufacturers make is that a user who physically owns a device should have full control over it. 
Thus, in \hid platforms, each device takes charge of authorization checking and sending device binding requests. 
This allows a legitimate user to rebind a device with another account by physically resetting it. 
Note that the rebind operation unbinds the previous account automatically and happens regardless of whether the device has already been unbound or not. 
This design directly leads to flaw F1.2, allowing an attacker to use a \pha device to remotely unbind the original user.

\paragraph{Device Reconnection.} 
Network congestion may cause random connection loss between a device and an IoT cloud. The IoT cloud mitigates the problem by allowing the device to re-login itself automatically. However, at the time of re-login, the cloud does not do any account-based authentication checking on all the platforms we have investigated except for SmartThings. This design gives raise to F3 which allows an attacker to remotely control the device without user awareness.

\paragraph{Cloud-Device State Inconsistency.} 
To avoid problematic state transitions, 
an IoT cloud should be aware of the status of the devices it manages.  
Unfortunately, this is very hard to achieve in practice. 
In the previous work, it has shown that 22 of 24 studied devices suffer
from design flaws that lead to state inconsistency~\cite{blindedconfused}.
For one thing, intermittent network conditions make it very difficult to keep the state of a device and the state of 
the IoT cloud synchronized at all time. 
For another, a user may reset a device by pushing the physical reset button when the device's Internet connection is lost.
As a result, the device binding information is cleared on the device but not in the cloud.
In all cases, the synchronization between the cloud and the device is broken. 
This causes flaws F1.1, F1.2, F1.3, and F2.

\vspace*{-2.5mm}
\subsection{Mitigation}
\vspace*{-2mm}
\label{sec:mitigation}

In this section, we propose several defensive design suggestions to secure smart home
platforms in the first place.
It should be noticed that adopting only a subset of our suggestions is not enough, because the flaws involved in the interactions are multi-faceted and tangled together (\eg~F1.3 and F3). 
Platform providers should consider all the potential security issues introduced by the interactions, including authentication, authorization and validity of working state machines.

\vspace*{-3mm}
\subsubsection{Strict Device Authentication}
\vspace*{-2mm} 
We have clearly shown that existing authentication is not adequate.
By violating authentication, a \pha device is indistinguishable from a victim device.
To ensure that every device an IoT cloud 
talks to is a genuine device, we suggest that
the manufacturers embed a unique client certificate into each device for high-end devices powered by Intel or ARM Cortex-A processors. In addition, the IoT cloud should always examine the client certificate before accepting any device request.
For resource-restricted devices powered by a microcontroller,
the manufacturers should embed a read-only random number into each device. 
On the cloud side, the cloud should always check whether the random number matches the other identity or legitimacy information.

Because device IDs are used by IoT clouds to identify a device,
we also suggest that platform providers 
retrofit the \id provisioning mechanism so that the attacker cannot easily obtain a valid \id.
Hard coding the device IDs is a bad practice because once a \id is leaked, the corresponding device 
becomes vulnerable forever. 
The \id of a device should be generated by the IoT cloud during registration, and the generation algorithm should use harder-to-guess information, such
as user ID/passwords, random numbers, etc.

\vspace*{-2.5mm} 
\subsubsection{Comprehensive Authorization Checking}
\vspace*{-1.5mm}
Compared with mobile-side commands,
we found that most IoT clouds do not enforce strict authorization checking of device-side commands and baselessly trust  arbitrarily connected devices.  
For \cid platforms, when a device talks to an IoT cloud, the user account information is absent on the device.
Thus, the IoT cloud directly accepts unauthorized logins (F3) or unbind (F4) commands. 
For \hid platforms, because the device takes charge of checking the binding relationship, the cloud skips performing further authorization checking on the requests from the device.

\label{sec:SmartThingsF3}
We suggest that both the device and the IoT cloud store and maintain the binding relationship
as well as perform authorization checking. 
Moreover, on the cloud side, the account-based authentication should be performed
on every device-side request, especially for critical operations such as device login. 
Samsung SmartThings follows this practice and thus is not vulnerable to flaw F3.
In SmartThings, devices must explicitly include user credentials for every login request. 
This additional credential checking prevents the target device from reconnecting to the cloud.




\vspace*{-2.5mm}
\subsubsection{Enforcing the Validity of State Transitions}
\vspace*{-1.5mm}



As revealed by our findings,
all the tested platforms failed to
enforce the validity of the involved state transitions.
In order to prevent the attacker from exploiting unexpected state transitions,  
smart home platforms should identify and formulate every legitimate interaction request as a 3-tuple in the form of (sender entity \& its state, the request message, receiver entity \& its state).  
In addition to checking every request, the 
sender entity should also verify if its current state allows the request to be sent out; and the receiver entity should verify if its current state is allowed to receive the request.
For instance, the IoT cloud shown in Figure 2 should only accept a device registration request when it stays in state 1. 
Furthermore, in order to prevent the three entities from staying out of the set of legitimate state combinations,  
the three entities should formally define and maintain their own state machines.  
In the meantime, the IoT cloud of a platform should synchronize the three entities so that they stay in a legitimate state combination.  
Finally, if an unrecoverable system error occurs,
the three entities should roll back to their initial states immediately.

\vspace*{-2.5mm}
\subsection{Malignant Commercial Competitions}
\vspace*{-1mm}
The discovered exploits could also be leveraged by
unscrupulous merchants in commercial competitions.

\paragraph{IP Theft.}
As mentioned in Section~\ref{sec:FT},
a company can steal a rival's firmware and
reverse-engineer it to steal proprietary IP.
This kind of behavior harms fair competition and hinders technology advancement.

\paragraph{Statistics Manipulation.}
By churning out hundreds of thousands of \pha devices,
a malicious company could rig the number of active devices in the market. This has two implications.
First, by increasing the market share of its own products, the company can present 
an eye-catching year-end report.
Second, by increasing the number of activated devices of its rival, 
its rival could be overcharged by the platform provider.
This is because some platform providers bill cooperative manufacturers 
based on the number of activated devices connected to their clouds.
A unscrupulous manufacturer can use \pha devices to
register a large number of non-existing devices under the name of its rivals,
causing financial loss to them.

\paragraph{User Experience Disruption.} 
Leveraging the illegal device occupation attack,
a unscrupulous manufacturer can potentially take over a large number of its rival's in-stock products.
When these products are sold, the consumers will have a terrible user experience.


\vspace*{-2mm}
\section{Related Work}
\label{RW}
\vspace*{-2mm}
We review the related work on smart home security from three perspectives:
device security, communication security and IoT application security.


\paragraph{Device Security.}
Device security research emphasizes the vulnerabilities of individual devices.
Ling \et~\cite{Ling2017Security} studied a smart plug system and revealed a weak authentication vulnerability.
After dissecting the behavior of several IoT devices such as Phillips Hue light bulbs and Nest smoke detectors,
Notra \et~\cite{Notra2014An} revealed that basic security mechanisms such as encryption, authentication and integrity checking are absent in these devices. 
Several currently available smart hubs were investigated in~\cite{hub} and~\cite{hub2}, and numerous security flaws were identified.
In contrast to analyzing individual devices, our study analyzes the complex interactions among the three entities engaged in a smart home platform. 




\paragraph{Communication Security.}
Communication security research emphasizes the security and privacy issues in 
smart home communication protocols such as BLE, ZigBee, and Z-Wave~\cite{HackingZWave,Agosta2016Cyber,Ronen2017IoT,Goyal2016Mind}.
Agosta \et ~\cite{Agosta2016Cyber} approached the security and privacy problems
involved in the key derivation algorithm adopted by the widespread Z-Wave home
automation protocol.
Ronen \et ~\cite{Ronen2017IoT} described a worm attack which has the
potential of massive spread by exploiting an implementation bug in the ZigBee Light Link protocol.
Researchers also demonstrated that attackers
can infer private in-home activities
by analyzing encrypted traffic from smart home devices~\cite{Apthorpe2017Spying} or by extracting features of connection-oriented application data unit exchanges~\cite{HomeSnitch}. 
Instead of focusing on a particular algorithm or protocol, this study conducts comprehensive platform-wide vulnerability analysis. 



\paragraph{IoT Application Security.}
Recently, increasing numbers of researchers have paid their attention to smart home platforms,
but they usually focus only on in-cloud IoT applications (i.e. home automation applications). 
For instance,
Fernandes \et~\cite{Fernandes2016Security} revealed that
the capabilities implemented in the SmartThings IoT application programming framework
are too coarse-grained, which allows malicious third-party
IoT applications to compromise the SmartThings platform.
Celik \et~\cite{celik2018sensitive}
proposed \textit{SAINT}, a static taint
analysis tool to find sensitive data flows
in IoT applications.
The same authors further studied whether an IoT application and its environment
adhere to functional safety properties.
They found that 9 out of 65 SmartThings apps violate 10 out of 35 properties~\cite{celik2018soteria}. 
Kafle \et~\cite{Kafle2018A} revealed the feasibility and severity (e.g., privilege escalation) of misuse of smart home routines. 
Moreover, Ding \et~\cite{ding2018safety} presented a tool named \textit{IoTMon} 
to discover risky interaction chains among IoT applications.
Our work focuses on the interactions between the participating entities engaged in a smart home platform, instead of between home automation applications.

\vspace*{-2mm}
\section{Conclusions}
\vspace*{-3mm} 
\label{conclusion}

Smart home technology is playing a more and more important role in our digital lives.
To seize a greater market share, smart home platform providers shorten the time-to-market by
reusing existing architectures and incorporating open-source projects without rigorous review (of the potential security and privacy issues).
In this work, we conducted an in-depth analysis of 
five widely-used smart home platforms, and found that the complex interactions among the participating entities (i.e., devices, IoT clouds, and mobile apps), though not being systematically investigated in the literature, are vulnerable to a spectrum of new attacks, including 
remote device substitution, remote device hijacking, remote device DoS, illegal device occupation, and firmware theft.   
The discovered vulnerabilities are applicable to multiple major smart home platforms, and cannot be amended via simple security patches.
Accordingly, we propose several defensive design suggestions 
to secure smart home platforms in the first place.

\vspace*{-2mm}
\section*{Acknowledgments}
\vspace*{-2.5mm}

We would like to thank our shepherd William Enck
and the anonymous reviewers for their helpful feedback.
Wei Zhou and Yuqing Zhang was support by National Key R\&D Program China (2016YFB0800700), National Natural Science Foundation of China (No.U1836210, No.61572460), Open Project Program of the State Key Laboratory of Information Security (2017-ZD-01) and in part by CSC scholarship.
Peng Liu was supported by ARO W911NF-13-1-0421 (MURI), 
NSF CNS-1505664, NSF CNS-1814679, and ARO W911NF-15-1-0576.
Le Guan was partially supported by JFSG from the University of Georgia Research Foundation, Inc.

{\footnotesize 
\bibliographystyle{plain}
\bibliography{bibs/iot}
}


\appendix
\renewcommand\thefigure{\arabic{figure}}
\renewcommand\thetable{\arabic{table}}

\FloatBarrier
\vspace*{-3mm}
\section{Legitimate 3-tuple State Combinations}
\vspace*{-4mm}
\label{sec:appendix1}
\setcounter{table}{0}
\setcounter{figure}{0} 
\begin{table}[h]
  \begin{center}
  \begin{adjustbox}{max width=\columnwidth}
  \begin{tabular}{c|ccc}
    \toprule
     &\tabincell{c}{\textbf{State of an IoT Cloud}}&
     \tabincell{c}{\textbf{State of a Device}}&
     \tabincell{c}{\textbf{State of a Mobile App}}\\
    \midrule
 \multirow{4}{*}{\tabincell{c}{\textbf{\cid}\\\textbf{Platform}}}
 &S1&S1 or S2&S1 or S2\\
 &S2&S3&S2\\
 &S3&S3&S3\\
 &S4&S4&S4
\cr\hline\hline
 \multirow{3}{*}{\tabincell{c}{\textbf{\hid}\\\textbf{Platform}}}&S2&S1 or S2&S1 or S2\\
&S3&S3&S3\\
&S4&S4&S4\\
  \bottomrule
\end{tabular}
\end{adjustbox}
\end{center}
\end{table}

\FloatBarrier
\vspace*{-3mm}
\section{Tested Devices and Applicable Attacks}
\vspace*{-4mm}
\label{sec:appendix4}
\begin{center}
\begin{table}[H]
  \begin{adjustbox}{width=\columnwidth}
  \begin{tabular}{c|cccc}
    \toprule 
     &\textbf{Tested Device}&
     \textbf{Device Model}&
     \textbf{Platform}&
     \textbf{Applicable Attacks}\\
    \midrule
    \multirow{5}{*}{\tabincell{c}{\\[4em]\textbf{\cid }\\\textbf{Platform}}}&\tabincell{c}{Mobile Remote\\HD Monitor}&\tabincell{c}{RIWYTH\\RW-821S-ALY}&Alink&\tabincell{c}{Remote Device Hijacking\\Remote Device Substitution\\Remote Device DoS\\Illegal Device Occupation}\\
    \cline{2-5}
    &WiFi Smart Adapter&\tabincell{c}{Philips\\ SPS9011A/93}&Alink&\tabincell{c}{Remote Device Hijacking\\Remote Device Substitution\\Remote Device DoS\\Illegal Device Occupation}\\
    \cline{2-5}
    &\tabincell{c}{Security\\Smart Lock}&\tabincell{c}{KAADAS\\ KDSLOCK001}&Alink&\tabincell{c}{Remote Device Hijacking\\Remote Device Substitution\\Remote Device DoS\\Illegal Device Occupation}\\
    \cline{2-5}
    &WiFi Smart Plug&\tabincell{c}{BULL\\ GN-Y2011}&Joylink&\tabincell{c}{Remote Device Substitution\\Illegal Device Occupation}\\
    \cline{2-5}
    &Smart Weighing Scale&ZK321J&Joylink&\tabincell{c}{Remote Device Substitution\\Illegal Device Occupation}
    \cr\hline\hline
    \multirow{6}{*}{\tabincell{c}{\\[3.5em]\textbf{\hid }\\\textbf{Platform}}}
    &WiFi Plug&\tabincell{c}{TP-Link\\ HS100}&KASA&\tabincell{c}{Remote Device Hijacking\\Remote Device Substitution\\Remote Device DoS}\\
    \cline{2-5}
    &WiFi LED Bulb&\tabincell{c}{TP-Link\\ LB110}&KASA&\tabincell{c}{Remote Device Hijacking\\Remote Device Substitution\\Remote Device DoS}\\
    \cline{2-5}
    &Smart Gateway&\tabincell{c}{XiaoMi\\ Multifunctional Gateway}&MIJIA&\tabincell{c}{Remote Device Hijacking\\Remote Device Substitution\\Remote Device DoS}\\
    \cline{2-5}
    &Smoke Alarm&\tabincell{c}{XiaoMi\\Fire Alarm Detector}&MIJIA&\tabincell{c}{Remote Device Hijacking\\Remote Device Substitution\\Remote Device DoS}\\
    \cline{2-5}   
    &Cleaning Robot&\tabincell{c}{Samsung\\POWERbot R7040}&SmartThings&Remote Device DoS\\
    \cline{2-5}
    &Hub&\tabincell{c}{Samsung\\Hub 3rd Generation}&SmartThings&Remote Device DoS\\
    \bottomrule
  \end{tabular}
  \end{adjustbox}
\caption{Tested devices and applicable attacks}
\end{table}
\end{center}

\FloatBarrier
\vspace*{-2mm}
\section{Sequence Diagrams}
\vspace*{-1mm}
We show the complete sequence diagram of interactions with
concrete parameters for each of the studied platforms.
Note that some essential steps are omitted because they are irrelevant to our attack.
In each figure, a box corresponds to
one phase in the life-cycle of
a device (Section~\ref{sec:subattack}).
\label{sec:appendix3}
\begin{figure}[h]
\begin{center}
\includegraphics[width=\columnwidth]{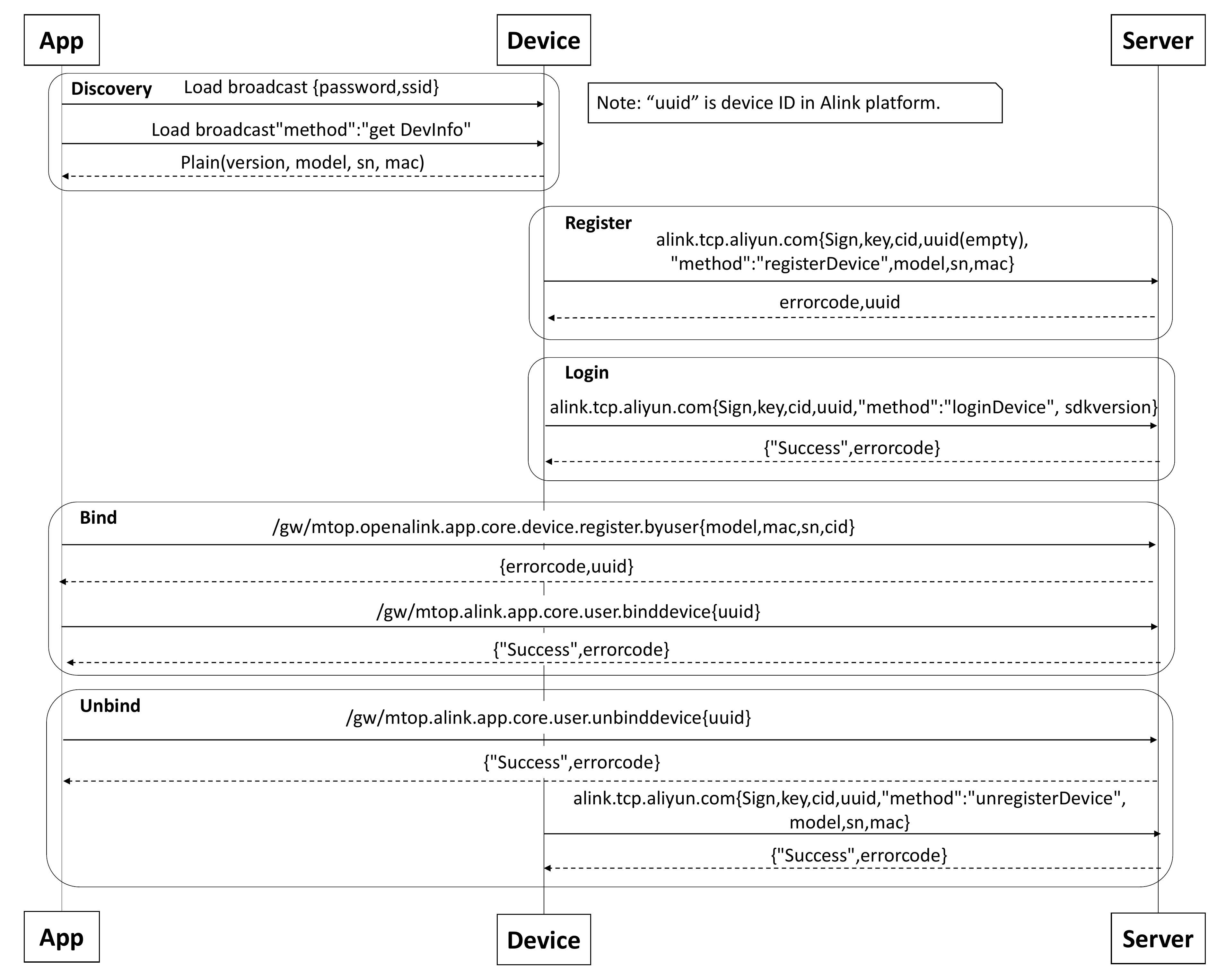}
\caption{Sequence Diagram of the Alink Platform}
\end{center}
\end{figure}
\vspace{-10em}

\begin{figure}[h]
\begin{center}
\includegraphics[width=\columnwidth]{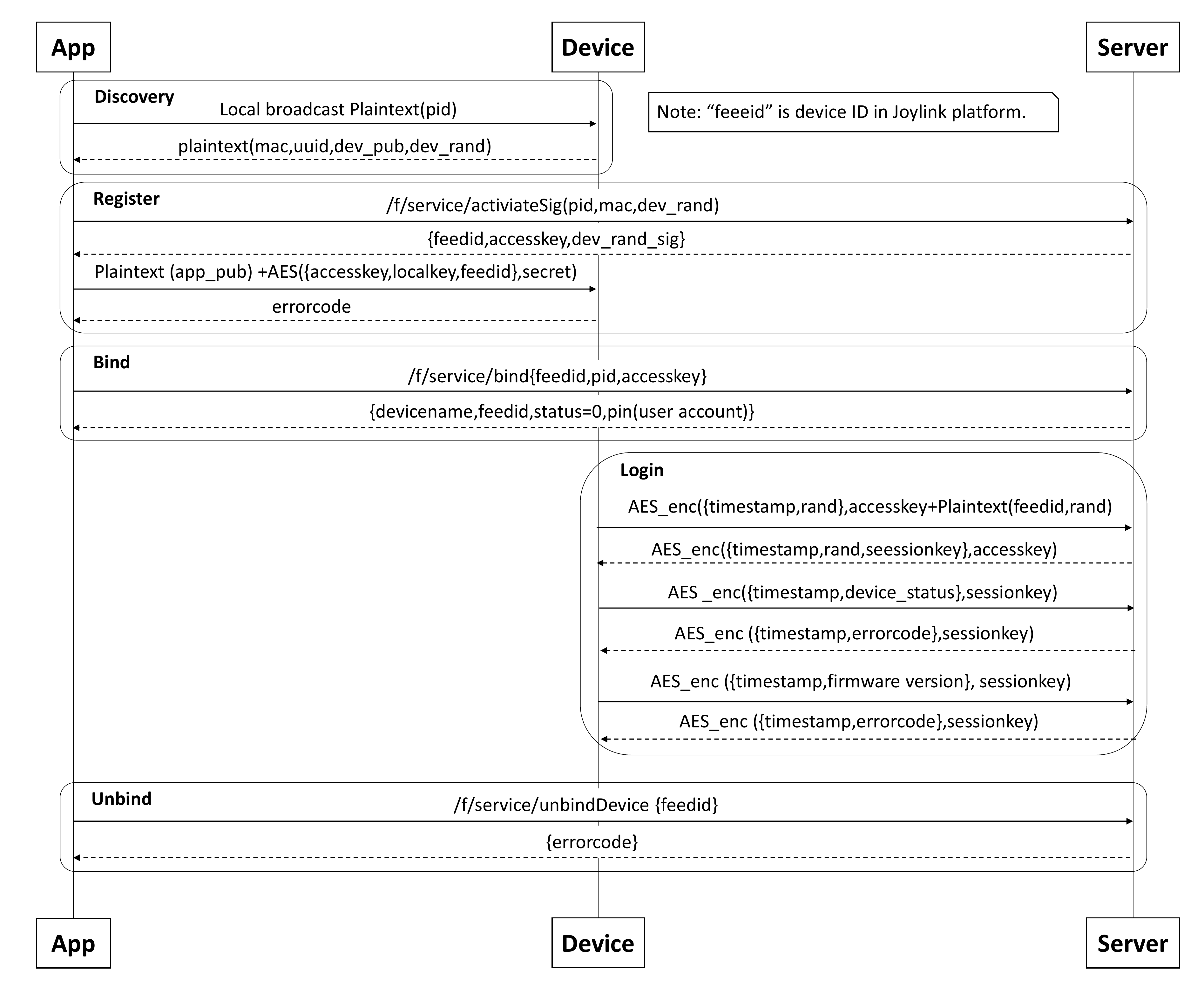}
\caption{Sequence Diagram of the Joylink Platform}
\end{center}
\end{figure}
\vspace{-10em}

\begin{figure}[h]
\begin{center}
\includegraphics[width=\columnwidth]{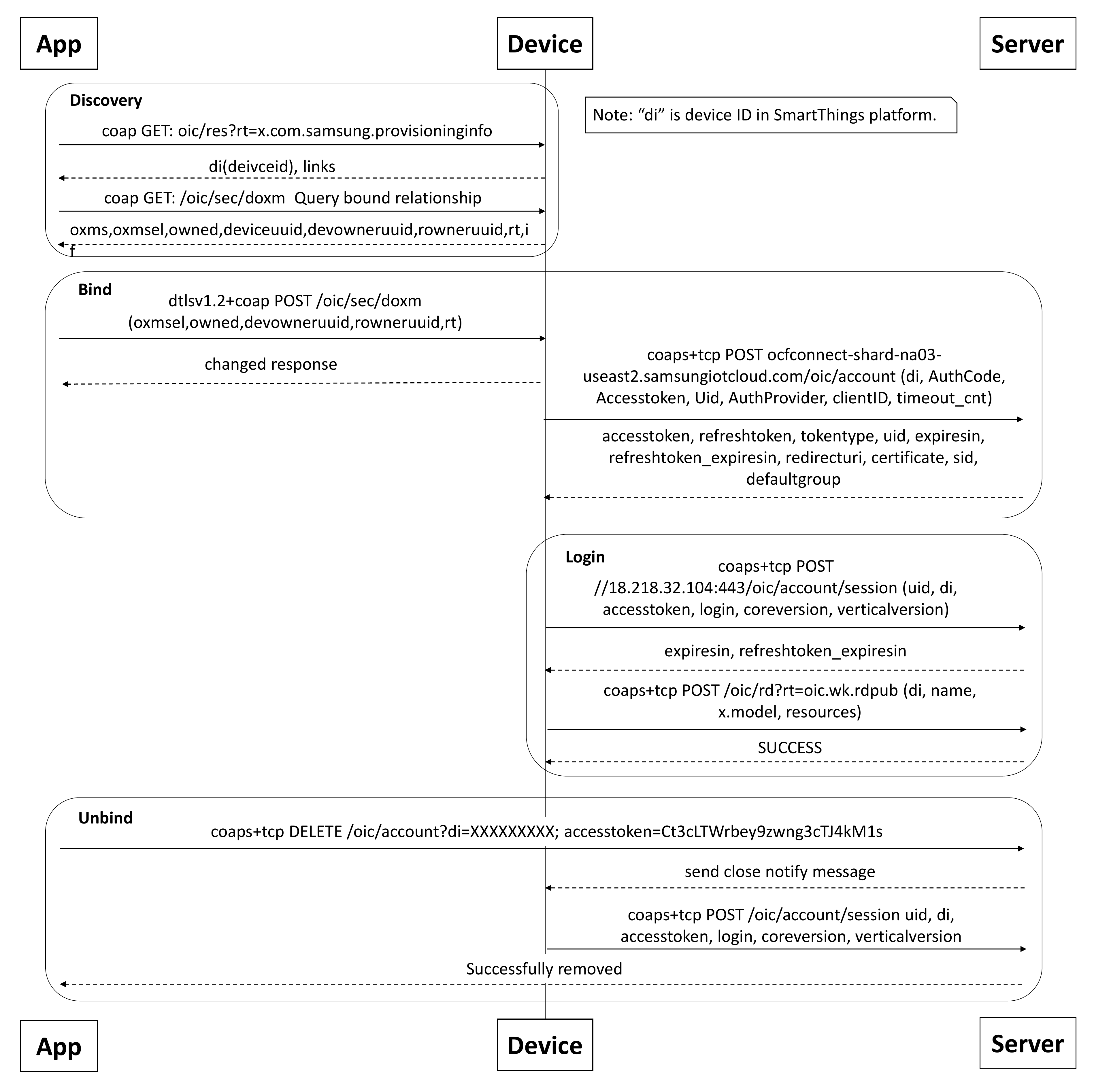}
\caption{Sequence Diagram of the SmartThings Platform}
\end{center}
\end{figure}
\vspace{-10em}

\begin{figure}[h]
\begin{center}
\includegraphics[width=\columnwidth]{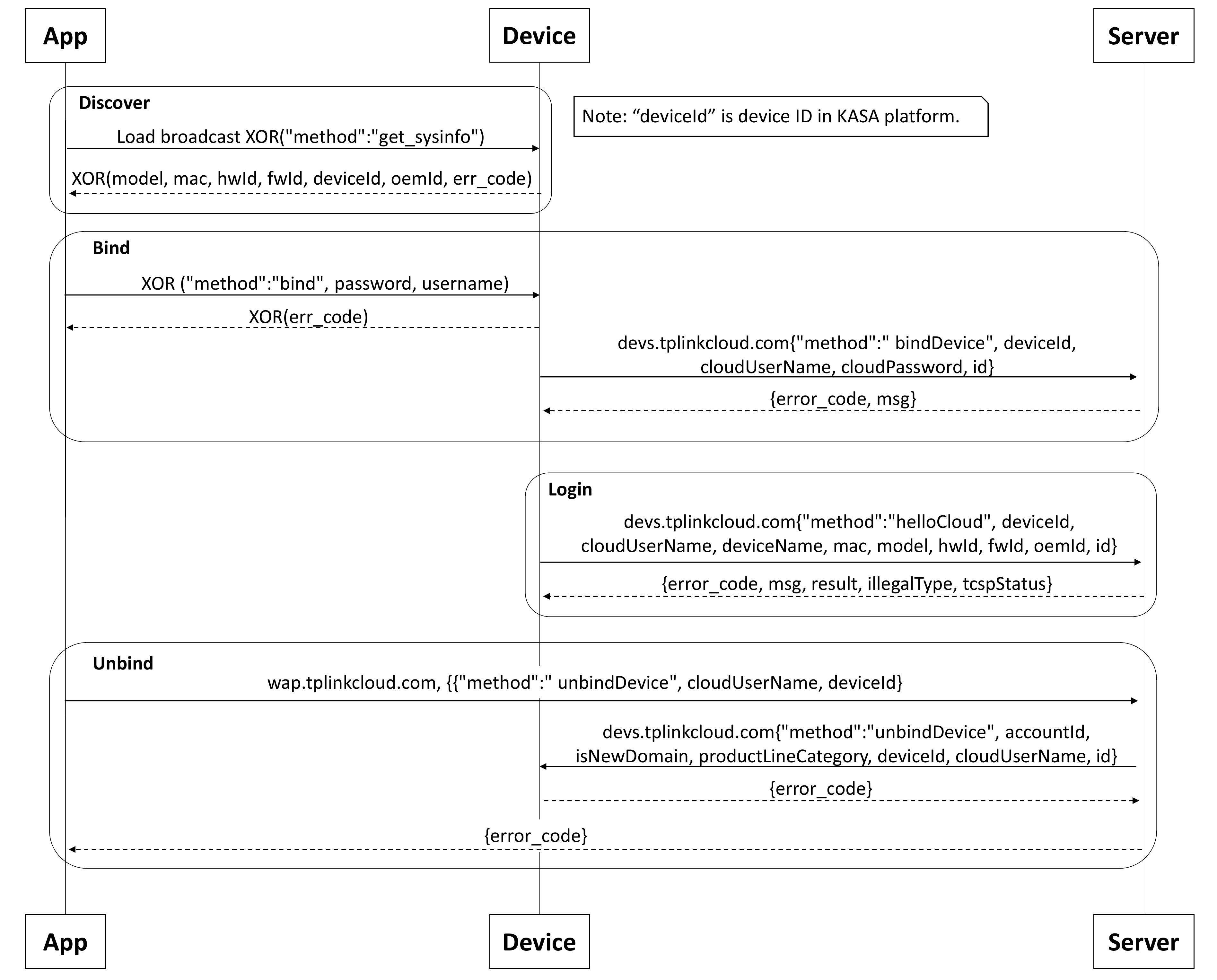}
\caption{Sequence Diagram of the KASA Platform}
\end{center}
\end{figure}

\begin{figure}[h]
\begin{center}
\includegraphics[width=\columnwidth]{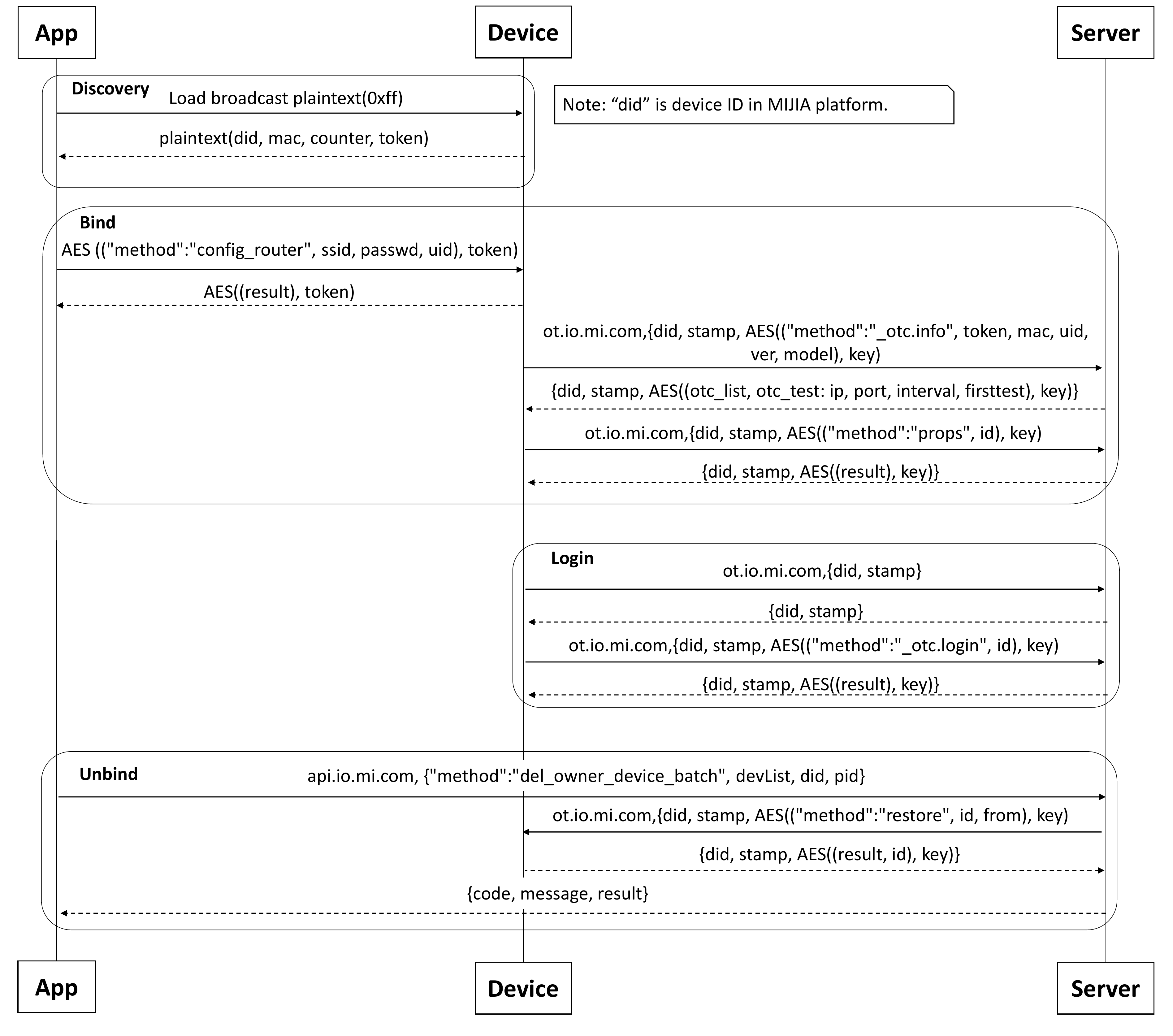}
\caption{Sequence Diagram of the MIJIA Platform}
\end{center}
\end{figure}


\end{document}